\documentclass{aa}
\usepackage{amssymb}
\usepackage{amsmath}
\usepackage{mathrsfs}
\usepackage{graphicx}
\usepackage{txfonts}
\usepackage[bookmarks, colorlinks, breaklinks]{hyperref}
\hypersetup{linkcolor=blue,citecolor=blue,filecolor=black,urlcolor=blue} 
\usepackage{natbib}
\bibpunct{(}{)}{;}{a}{}{,} 

\newcommand{\code}[1]{\texttt{#1}}
\newcommand{\micron}{\mbox{$\mu$m}}

\begin{document} 

\title{Benchmarking the Calculation of Stochastic Heating \\ and Emissivity of Dust Grains \\ in the Context of Radiative Transfer Simulations}
\titlerunning{Benchmark: Stochastic Heating of Dust Grains}

\author{Peter Camps \inst{\ref{ugent}}
  \and Karl Misselt \inst{\ref{steward}}
  \and Simone Bianchi \inst{\ref{arcetri}}
  \and Tuomas Lunttila \inst{\ref{chalmers}}
  \and Christophe Pinte \inst{\ref{umi},\ref{ipag}}
  \and Giovanni Natale \inst{\ref{preston}}
  \and \\ Mika Juvela \inst{\ref{helsinki}}
  \and Joerg Fischera \inst{\ref{maxplanck1}}
  \and Michael P. Fitzgerald \inst{\ref{losangeles}}
  \and Karl Gordon \inst{\ref{mary},\ref{ugent}}
  \and Maarten Baes \inst{\ref{ugent}}
  \and J\"urgen Steinacker \inst{\ref{ipag},\ref{maxplanck2}}
}
\authorrunning{P. Camps et al.}

\institute{Sterrenkundig Observatorium, Universiteit Gent, Krijgslaan 281, B-9000 Gent, Belgi\"e, \email{peter.camps@ugent.be} \label{ugent}
  \and Steward Observatory, University of Arizona, Tucson, AZ 85721-0065, USA \label{steward}
  \and INAF-Osservatorio Astrofisico di Arcetri, Largo Enrico Fermi 5, I-50125 Firenze, Italia \label{arcetri}
  \and Chalmers University of Technology, Department of Earth and Space Sciences, Onsala Space Observatory, SE-439 92 Onsala, Sweden \label{chalmers}
  \and UMI-FCA, CNRS-INSU, France (UMI 3386) and Universidad de Chile, Cerro Cal\'an, Santiago, Chile \label{umi}
  \and Univ. Grenoble Alpes, IPAG, F-38000 Grenoble, France \\ CNRS, IPAG, F-38000 Grenoble, France \label{ipag}
  \and University of Central Lancashire, Jeremiah Horrocks Institute, Leighton building, Preston, PR1 2HE, UK \label{preston}
  \and Department of Physics, PO Box 64, University of Helsinki, 00014 Helsinki, Finland  \label{helsinki}
  \and Max Planck Institut f\"ur Kernphysik, Saupfercheckweg 1, D-69117 Heidelberg, Germany \label{maxplanck1}
  \and Department of Physics and Astronomy, 430 Portola Plaza, University of California, Los Angeles, CA 90095-1547, USA \label{losangeles}
  \and Space Telescope Science Institute, 3700 San Martin Drive, Baltimore, Maryland 21218, USA \label{mary}
  \and Max-Planck-Institut f\"ur Astronomie, K\"onigstuhl 17, D-69117 Heidelberg, Germany \label{maxplanck2}
}
\date{\today}

\abstract
   {Thermal emission by stochastically heated dust grains (SHGs) plays an important role in the radiative transfer (RT) problem for a dusty medium. It is therefore essential to verify that RT codes properly calculate the dust emission before studying the effects of spatial distribution and other model parameters on the simulated observables.}
   {We define an appropriate problem for benchmarking dust emissivity calculations in the context of RT simulations, specifically including the emission from SHGs. Our aim is to provide a self-contained guide for implementors of such functionality, and to offer insights in the effects of the various approximations and heuristics implemented by the participating codes to accelerate the calculations.}
   {The benchmark problem definition includes the optical and calorimetric material properties, and the grain size distributions, for a typical astronomical dust mixture with silicate, graphite and PAH components; a series of analytically defined radiation fields to which the dust population is to be exposed; and instructions for the desired output. We process this problem using six RT codes participating in this benchmark effort, and compare the results to a reference solution computed with the publicly available dust emission code DustEM.}
   {The participating codes implement different heuristics to keep the calculation time at an acceptable level. We study the effects of these mechanisms on the calculated solutions, and report on the level of (dis)agreement between the participating codes. For all but the most extreme input fields, we find agreement within 10\% across the important wavelength range $3\,\micron\le\lambda\le1000\,\micron$.}
   {We conclude that the relevant modules in RT codes can and do produce fairly consistent results for the emissivity spectra of SHGs. This work can serve as a reference for implementors of dust RT codes, and it will pave the way for a more extensive benchmark effort focusing on the RT aspects of the various codes.}


\maketitle


\section{Introduction\label{sec:introduction}}

Dust substantially affects the radiation emerging from many astrophysical systems. To study the three-dimensional structure of these systems, it is often useful to numerically simulate the transport of radiation through a model that includes a dusty medium with appropriate characteristics. Many authors have described radiative transfer (RT) codes designed to tackle this problem; for an overview see for example \citet{2011BASI...39..101W} and \citet{2013ARAA..51...63S}. In multi-wavelength studies, thermal emission by the dust plays an important role. While larger dust grains can often be assumed to be in local thermal equilibrium (LTE) with the surrounding radiation field, this assumption does not typically holds for very small grains (VSGs) or for polycyclic aromatic hydrocarbon molecules (PAHs). The absorption of a single optical photon substantially boosts the internal energy of such a small collection of atoms, causing its emission spectrum to vary over time \citep[see, e.g.,][]{1984ApJ...277..623S,1986ApJ...302..363D,1988ApJ...330..964B,2000ApJ...532L..21H,2007ApJ...663..866D,2007ApJ...656..770S}. We use the term \emph{stochastically heated grains} (SHGs) to collectively indicate dust grains and PAH molecules that can not be assumed to be in LTE with the radiation field.

Since SHGs spend a significant amount of time at much higher energy levels than if they would be in LTE, they emit at shorter wavelengths, which can have an important effect on the observed spectrum. It is therefore essential to verify that RT codes properly calculate this emission before studying the effects of spatial distribution and other model parameters on the simulated observables. In this paper we present a benchmark problem for this purpose, including a dust model, a series of input radiation fields, and a reference dust emission spectrum for each input field. The dust model described here has been designed for use in the TRUST benchmarks\footnote{TRUST benchmarks: \url{http://ipag.osug.fr/RT13/RTTRUST/}} (Transport of Radiation through a dUSTy medium), which test the actual RT aspect of various codes.

A typical 3D RT simulation calculates the dust emission spectra for millions of dust cells (or at least for many thousands of library items that are representative of the cells). When calculating the emission spectrum for the dust grain population in a particular cell, the first task is to determine the temperature probability distribution of the grains, given the grain sizes and chemical compositions, and given the radiation field in the cell. This is also the computationally most demanding part of the calculation. In the current epoch, performing a full treatment of vibrational quantum modes for the dust grains in each cell is computationally prohibitive. In practice, RT codes use the continuous cooling assumption, which was shown to provide a good approximation in areas relevant for RT through the interstellar medium \citep{2001ApJ...551..807D}. 

The reference solutions in this paper were produced with the public version of the DustEM code \citep{2011AA...525A.103C}. DustEM determines the grain temperature distribution by iteratively solving an integral equation \citep{1986AA...160..295D}. We compare the reference solutions with the dust emission spectra calculated by six distinct RT codes. These codes determine the grain temperature distribution by solving a set of linear equations \citep{1989ApJ...345..230G}. While this method is inherently faster \citep{1989ApJ...345..230G}, it still becomes very expensive when the grains are in LTE with the radiation field. To avoid this problem, care must be taken to properly transition the calculation from the stochastic to the equilibrium regime.

More generally, the need for fast dust emission calculations has prompted the authors of RT simulation codes to implement various acceleration techniques, discretization choices, approximations, and heuristics. We study the effects of these mechanisms on the calculated solutions, and quantify the level of (dis)agreement between the participating codes, with the objective to help inform the interpretation of RT simulation results that include SHG dust emission calculations of the type presented here.

The information provided in this paper and in the accompanying data files\footnote{\label{fn-website}SHG benchmark data: \url{http://www.shg.ugent.be}} is self-contained. Readers can implement the code to calculate the SHG emission, set up the benchmark tests, and verify the results, solely based on the information provided here, without referring to other sources.

In Sect.\,\ref{sec:model} and Sect.\,\ref{sec:fields} we define the dust model and the input radiation fields used in the benchmark problem. Section \ref{sec:emission} presents the linear equation method used by the RT codes represented in this paper to calculate the SHG emission spectra, and Sect.\,\ref{sec:codes} briefly introduces each of these codes. In Sect.\,\ref{sec:results} we compare the results produced by the RT codes with the reference solution, and we discuss the differences between the methods used by the various codes and how they influence the results. Finally, in Sect.\,\ref{sec:conclusions} we summarize and conclude.

For ease of reference, Table \ref{tbl:symbols} lists the symbols used in this paper for various physical quantities, with the corresponding SI units.

\begin{table*}
\caption{Symbols used in this paper for various physical quantities, with the corresponding SI units.}
\label{tbl:symbols}
{\renewcommand{\arraystretch}{1.15}
\begin{tabular}{l l l}
\hline\hline
Symbol & Units & Description \\
\hline
$\lambda$ & $\mathrm{m}$ & Wavelength \\
$s$ & $\mathrm{m}$ & Distance along a path \\
$V$ & $\mathrm{m}^3$ & Volume \\
$M$ & $\mathrm{kg}$ & Dust mass \\
$T$ & K & Temperature \\
$\tau$ & s & Interaction timescale \\
\hline
$\rho$ & $\mathrm{kg}\,\mathrm{m}^{-3}$ & Dust mass density \\
$\sigma^{\mathrm{abs},\mathrm{sca},\mathrm{ext}}$ & $\mathrm{m}^2$ & Cross section (absorption, scattering or extinction) \\
$\mathcal{N}_\mathrm{H}$ & $\mathrm{H}$ & Number of hydrogen atoms \\
$n_\mathrm{H}=\mathcal{N}_\mathrm{H}/V$ & $\mathrm{H}\,\mathrm{m}^{-3}$ & Hydrogen atom number density \\
$\mu=M/\mathcal{N}_\mathrm{H}$ & $\mathrm{kg}\,\mathrm{H}^{-1}$ & Dust mass per hydrogen atom \\
$\varsigma=\sigma/\mathcal{N}_\mathrm{H}$ & $\mathrm{m}^2\,\mathrm{H}^{-1}$ & Cross section per hydrogen atom (absorption, scattering or extinction) \\
$\kappa=\sigma/M=\varsigma/\mu$ & $\mathrm{m}^2\,\mathrm{kg}^{-1}$ & Mass coefficient (absorption, scattering or extinction) \\
\hline
$a$ & $\mathrm{m}$ & Dust grain size \\
$\mathcal{N}_\mathrm{D}$ & $\mathrm{1}$ & Number of dust grains \\
$n_\mathrm{D}=\mathcal{N}_\mathrm{D}/V$ & $\mathrm{m}^{-3}$ & Dust grain number density \\
$\Omega(a)=(\frac{\mathrm{d}n_\mathrm{D}}{\mathrm{d}a})/n_\mathrm{H}$ & $\mathrm{m}^{-1}\,\mathrm{H}^{-1}$ & Dust grain size distribution per hydrogen atom \\
$Q^{\mathrm{abs},\mathrm{sca},\mathrm{ext}}$ & $1$ & Efficiency (absorption, scattering or extinction) \\
$\rho^\mathrm{bulk}$ & $\mathrm{kg}\,\mathrm{m}^{-3}$ & Bulk mass density of grain material\\
$c$ & $\mathrm{J}\,\mathrm{K}^{-1}\,\mathrm{kg}^{-1}$ & Specific heat capacity of grain material \\
$h$ & $\mathrm{J}\,\mathrm{kg}^{-1}$ & Specific enthalpy (internal energy per unit mass) \\
$H$ & $\mathrm{J}$ & Enthalpy (internal energy) of a dust grain \\
\hline
$J$ & $\mathrm{W}\,\mathrm{m}^{-3}\,\mathrm{sr}^{-1}$ & Mean spectral radiance (intensity of radiation field) \\
$B$ & $\mathrm{W}\,\mathrm{m}^{-3}\,\mathrm{sr}^{-1}$ & Black-body spectral radiance (Planck's law) \\
$U$ & $1$ & Radiation field strength relative to solar neighborhood \\
$j$ & $\mathrm{W}\,\mathrm{m}^{-1}\,\mathrm{sr}^{-1}$ & Spectral dust emission \\
$\varepsilon = j/\mathcal{N}_\mathrm{H}$ & $\mathrm{W}\,\mathrm{m}^{-1}\,\mathrm{sr}^{-1}\,\mathrm{H}^{-1}$ & Spectral dust emission per hydrogen atom \\
\hline
\end{tabular}}
\end{table*}


\section{Dust model\label{sec:model}}

The exact choice of dust grain model \citep[e.g.,][]{2001ApJ...548..296W,2004ApJS..152..211Z,2013A&A...558A..62J} is not critical for benchmark purposes, as long as all codes employ the same model. Specifically, our choices do not imply a preference for or an endorsement of a particular model. Still, to properly evaluate the results of our calculations and the effects of approximations in the context of real-world RT simulations of astrophysical systems, it is important to use grain compositions and size distributions that are in agreement with observational constraints, rather than defining arbitrary synthetic grain properties. With this in mind, we have elected to utilize the simple \mbox{BARE-GR-S} model of \citet{2004ApJS..152..211Z}. This model uses a mixture of spherical, single composition (BARE) graphitic (GR) and silicate (S) dust grains.  The graphitic grains include both graphite grains as well as astronomical PAH molecules. The relative populations of the three components and their size distribution has been optimized to reproduce observations characteristic of the diffuse Milky Way interstellar medium with $R_V=3.1$. The observational constraints on the model include the shape of the wavelength dependent extinction, the infrared emission, and the abundance of elements locked up in the solid phase.

Assuming that the scattering process is elastic, so that the internal energy of the interacting dust grain remains unaffected, scattering is irrelevant for the benchmark problem presented in this paper. Thus the subsequent discussion does not focus on the scattering properties of the dust model.

\subsection{Optical grain properties}

For both the graphite and silicate components, optical properties were computed directly from the dielectric functions (refractive indices) of the bulk material using a Mie code originally provided by Viktor Zubko with some small modifications. The optical properties include absorption efficiencies $Q^\mathrm{abs}_{k}(a,\lambda)$ as a function of composition $k$, grain size $a$, and wavelength $\lambda$. 

While the dielectric functions are in general functions of both temperature and size \citep{1984ApJ...285...89D}, to minimize free parameters and in keeping with essentially all astronomical applications of grain properties, we adopt a single set of dielectric functions at a specific temperature and size for each component. The graphitic indices were taken from \citet{2003ApJ...598.1026D} for 0.1\,\micron\ grains at 20\,K. Calculations were carried out for both parallel and perpendicular orientations relative to the basal plane and combined using the 1/3-2/3 approximation \citep{1988ApJ...333..848D}. For the silicate grains, the dielectric functions for smoothed astronomical silicates (0.1\,\micron) \citep{1993ApJ...402..441L,2001ApJ...548..296W} were used in the Mie calculations. With these sets of dielectric functions, assuming sperical grains, the efficiencies can be calculated for each component as a function of size and wavelength. For each component, the Mie calculation was carried out for 121 logarithmically spaced sizes between 0.00035\,\micron\ and 100\,\micron.  For each size, optical properties were computed at 1201 wavelength points logarithmically spaced between 0.001\,\micron\ and 10000\,\micron. The efficiencies thus derived can be cast as either cross sections or mass coefficients:
\begin{align}
\sigma^\mathrm{abs,sca}_{k} (a,\lambda) &= \pi a^{2} Q^\mathrm{abs,sca}_{k}(a,\lambda) \label{eq:single-cross} \\
\kappa^\mathrm{abs,sca}_{k}(a,\lambda)  &= \frac{3 Q^\mathrm{abs,sca}_{k}(a,\lambda)}{4 a \rho_{k}}.
\end{align}

As there are no refractive indices for the class of materials commonly referred to as astronomical PAH, the optical properties for these materials are simply constructed so as to reproduce the available observations. Therefore, the PAH cross sections were computed following \citet{2007ApJ...657..810D}. The utilization of the \citet{2007ApJ...657..810D} formulation for the cross sections represents a deviation from a pure \citet{2004ApJS..152..211Z}
\mbox{BARE-GR-S} model as \citet{2004ApJS..152..211Z} utilized the \citet{2001ApJ...554..778L} PAH cross sections.

In \citet{2007ApJ...657..810D}, the PAH cross sections were updated to reflect knowledge gained regarding the shape of the PAH emission spectrum from the wealth of Spitzer observations available. Also of note, the PAH component in the model defined here consist of pure neutral PAHs; no attempt was made to account for the varying ionization fraction of a PAH molecule as a function of effective PAH size and radiation field. The PAH optical properties were computed on the same wavelength grid as the graphite and silicate components of the model. However, the size grid was of course altered; properties were computed for 28 logarithmically spaced sizes between 0.00035\,\micron\ and 0.006\,\micron.

\subsection{Grain size distributions\label{sec:sdist}}

\begin{table*}
\caption{Parameters for the analytical approximation to the BARE-GR-S size distribution defined in Eqs.\,\ref{eq:form-sdist} through \ref{eq:full-sdist} and the bulk densities for the three components in our dust model. With these parameter values, the grain size $a$ substituted in the equations must be expressed in \micron\ and the resulting size distribution $\Omega_{k}(a)$ is expressed in number of dust grains per hydrogen atom and per \micron.}
\label{tbl:baregrs}
\label{tbl:rhobulk}
{\renewcommand{\arraystretch}{1.25}
\begin{tabular}{c c c c c}
\hline\hline
Parameter & Units & PAH & Graphite & Silicate \\
\hline
$a^\mathrm{min}$ & \micron & 0.00035      & 0.00035       & 0.00035             \\
$a^\mathrm{max}$ & \micron & 0.005         & 0.33            & 0.37                   \\
\hline
$A$  & $\micron^{-1}\,\mathrm{H}^{-1}$    & 2.227433$\times 10^{-7}$ & 1.905816$\times 10^{-7}$ & 1.471288$\times 10^{-7}$ \\
$c_{0}$  & 1           & -8.02895                 & -9.86000                 & -8.47091                 \\
$b_{0}$  & 1           & -3.45764                 & -5.02082                 & -3.68708                 \\
$b_{1}$  & 1           & 1.18396$\times 10^{3}$   & 5.81215$\times 10^{-3}$  & 2.37316$\times 10^{-5}$    \\
$a_{1}$  & \micron  &  1                         & 0.415861                 & 7.64943$\times 10^{-3}$  \\
$m_{1}$ & 1           & -8.20551                 & 4.63229                  & 22.5489                  \\
$b_{2}$  & 1           & 0                 & 0                  & 0                  \\
$a_{2}$  & \micron  & 1                 & 1                  & 1                   \\
$m_{2}$ & 1           & 1                  & 1                  & 1                  \\
$b_{3}$  & 1           & 1.0$\times 10^{24}$      & 1.12502$\times 10^{3}$   & 2.96128$\times 10^{3}$   \\
$a_{3}$  & \micron  & -5.29496$\times 10^{-3}$ & 0.160344             & 0.480229                 \\ 
$m_{3}$ & 1           & 12.0146                  & 3.69897                  & 12.1717                  \\
$b_{4}$  & 1           &0                  & 1.12602$\times 10^{3}$   & 0                  \\
$a_{4}$  & \micron  & 0                  & 0.160501                 & 0                  \\
$m_{4}$ & 1           & 1                 & 3.69967                  & 1                  \\
\hline
$\rho^\mathrm{bulk}$ & $\mathrm{kg}\,\mathrm{m}^{-3}$ & 2240 & 2240 & 3500 \\
\hline
\end{tabular}}
\end{table*}

The relative contribution of each material in a dust model is set by the grain size distribution for that particular material. The shape of the size distributions in the \mbox{BARE-GR-S} model of \citet{2004ApJS..152..211Z} matches observations of the interstellar medium in the solar neighborhood, including the amount of refractory material available, as well as the wavelength dependence of the Milky Way diffuse ($R_V = 3.1$) extinction and emission spectrum. \citet{2004ApJS..152..211Z} provide convenient analytic functional approximations to the size distributions for each component of the model, which take the form 
\begin{equation}
\begin{split}
\log O(a) = c_{0} + b_{0}\log\left(\frac{a}{\micron}\right) - b_{1}\left|\log\left(\frac{a}{a_1}\right)\right|^{m_{1}} - b_{2}\left|\log\left(\frac{a}{a_2}\right)\right|^{m_{2}} \\ 
 - b_{3}\left| \frac{a-a_{3}}{\micron}\right|^{m_{3}} - b_{4}\left| \frac{a-a_{4}}{\micron}\right|^{m_{4}} 
\end{split}
\label{eq:form-sdist}
\end{equation}
where $a$ is the grain size and $a_i,b_i,c_i,m_i$ represent constant parameters. In the interest of clarity, the notation in Eq.\,\ref{eq:form-sdist} omits a subscript $k$ indicating the type of material being referenced. For the \mbox{BARE-GR-S} model, we require three sets of parameters, one for each of the materials comprising the dust model. The appropriate values can be found in Table 7 of \citet{2004ApJS..152..211Z}, and are reproduced here in Table \ref{tbl:baregrs}.

By construction, Eq. \ref{eq:form-sdist} has been normalized such that 
\begin{equation}
\int_{a^\mathrm{min}_k}^{a^\mathrm{max}_k} O_{k}(a) \,\mathrm{d}a = 1
\end{equation}
where $a^\mathrm{min}_k$ and $a^\mathrm{max}_k$ are the minimum and maximum sizes over which the size distribution for component $k$ is defined. The complete size distribution function is then given by 
\begin{equation}
\Omega_{k}(a) = A_{k} \, O_{k}(a) 
\label{eq:full-sdist}
\end{equation}
where $A_{k}$ is the overall normalization of component $k$, as listed in Table \ref{tbl:baregrs}.

With this definition of the grain size distribution, the total number of dust grains per hydrogen atom and the total dust mass per hydrogen atom are given by:
\begin{align}
    \mathcal{N}_\mathrm{D}/\mathcal{N}_\mathrm{H} &= \sum_{k}\int_{a^\mathrm{min}_k}^{a^\mathrm{max}_k} \Omega_{k}(a) \,\mathrm{d}a \\
    \mu = M/\mathcal{N}_\mathrm{H} &= \sum_{k}\int_{a^\mathrm{min}_k}^{a^\mathrm{max}_k} \frac{4}{3}\pi a^{3}\rho^\mathrm{bulk}_{k}  \Omega_{k}(a) \,\mathrm{d}a.
    \label{eq:dustmass}
\end{align}

We also provide the grain size distributions defined by Eqs.\,\ref{eq:form-sdist} and \ref{eq:full-sdist} and Table \ref{tbl:baregrs} in tabulated form (see Sect.\,\ref{sec:model-data}), on a size grid with 24, 62, and 63 samples for the PAH, graphite, and silicate component, respectively.

\subsection{Calorimetric grain properties}

The internal energy of a dust grain of composition $k$ and its temperature are related via the heat capacity of the grain material:
\begin{align}
H_k(a,T) &= \frac{4}{3}\pi a^{3}\rho^\mathrm{bulk}_{k} h_k(T) \label{eq:enthalpy} \\
h_k(T) &= \int_0^T c_k(T')\,\mathrm{d}T' 
\end{align}
where $H(a,T)$ is the internal energy (enthalpy) of a grain with size $a$ at temperature $T$; $h(T)$ is the specific enthalpy (per unit mass); $c(T)$ is the specific heat capacity; and $\rho^\mathrm{bulk}$ is the bulk density of the grain material.

We elected to use the heat capacity functions proposed in \citet{2001ApJ...551..807D} and \cite{2001ApJ...554..778L}. To avoid subtle differences between implementations, we provide this information in tabulated form (see Sect.\,\ref{sec:model-data}) rather than expecting each code to implement the equations. One table describes the graphitic components (PAH molecules and graphite grains) and another table describes the silicate component. Each table lists the specific enthalpy and the specific heat capacity at 1000 logarithmically spaced temperature points ranging from 1\,K to 2500\,K. 

Finally, the bulk densities for the three components in our dust model are listed in Table \ref{tbl:rhobulk}.

\subsection {Data files\label{sec:model-data}}

The data files defining the dust properties described above can be downloaded from the web site indicated in footnote \ref{fn-website}. They are contained in the \code{DustModel} directory, which is organized in the following subdirectories:
\begin{itemize}
\item \code{GrainInputs}: optical and calorimetric properties for each of the three grain types $k$. Optical properties include the absorption efficiencies $Q_k^{\mathrm{abs}}(a,\lambda)$ tabulated on a grid of 1201 wavelengths $\lambda$ and 28 (for PAH) or 121 (for graphite and silicate) grain sizes $a$. Calorimetric properties include the specific enthalpy $h(T)$ and the specific heat capacity $c(T)$ of the grain material, tabulated on a grid of 1000 temperatures $T$. The calorimetry data for graphitic grains should be used for PAH molecules as well. 
\item \code{SizeInputs}: tabulated grain size distributions $\Omega_k(a)$ for each grain type $k$, on a size grid with 24, 62, and 63 samples for the PAH, graphite, and silicate component, respectively. Implementations may choose to compute the size distribution from the functional form defined in Sect.\,\ref{sec:sdist}, or to load the tabulated data.
\item \code{EffectiveGrain}: size-integrated values of the optical properties. This information is not needed for calculating dust emission.
\item \code{ScatMatrix}: scattering matrix elements. This information is not needed for calculating dust emission.
\end{itemize}


\section{Radiation fields\label{sec:fields}}

\subsection{Basic definitions\label{sec:rad-defs}}

The spectral radiation for a black body in thermal equilibrium at temperature $T$ in function of wavelength $\lambda$ is given by the Planck function
\begin{equation}
B(\lambda,T)=\frac{2hc^{2}}{\lambda^{5}}\frac{1}{\exp(\frac{hc}{\lambda kT})-1}
\label{eq:planck}
\end{equation}
where $h$ denotes the Planck constant, $c$ the speed of light in vacuum, and $k$ the Boltzmann constant.

We define the solar neighborhood interstellar radiation field (ISRF) given in table A3 of \citet{1983A&A...128..212M} through the following functional form inspired by (but not identical to\footnote{The \citet{2001ApJS..134..263W} equation is formulated in function of frequency rather than wavelength, and the dilution factor for the 4000\,K black body listed in their Table 1 is not adjusted to the value specified in Sect. 2.1 of \citet{1983A&A...128..212M}.}) Eq. 31 in \citet{2001ApJS..134..263W}:
\begin{equation}
J^\mathrm{Mat}(\lambda)=\begin{cases}
0 \\ \hspace{4.2cm}  \lambda<0.0912\,\micron\\
3069\,\mathrm{W}/\mathrm{m}^{3}/\mathrm{sr}\times(\lambda/\micron)^{3.4172} \\ \hfill 0.0912\,\micron\leq\lambda<0.110\,\micron\\
1.627\,\mathrm{W}/\mathrm{m}^{3}/\mathrm{sr} \\ \hfill 0.110\,\micron\leq\lambda<0.134\,\micron\\
0.0566\,\mathrm{W}/\mathrm{m}^{3}/\mathrm{sr}\times(\lambda/\micron)^{-1.6678} \\ \hfill 0.134\,\micron\leq\lambda<0.250\,\micron\\
\begin{split}10^{-14}\,B(\lambda,7500\,\mbox{K})\,+\,10^{-13}\,B(\lambda,4000\,\mbox{K})\\ +\,4\times10^{-13}\,B(\lambda,3000\,\mbox{K})\end{split} \\ \hfill 0.250\,\micron\leq\lambda
\end{cases}
\label{eq:mathis}
\end{equation}
Note that the recipes in the other papers prescribe the total radiation field $4\pi J^{\mathrm{Mat}}$, whereas we prescribe the mean radiation field $J^{\mathrm{Mat}}$. Based on this reference field, we define the strength $U$ of an arbitrary radiation field $J(\lambda)$ as
\begin{equation}
U=\left. \int_{0}^{\infty}J(\lambda)\,\mathrm{d}\lambda \;\; \middle/ \; \int_{0}^{\infty}J^{\mathrm{Mat}}(\lambda)\,\mathrm{d}\lambda \right.
\end{equation}

\subsection{Benchmark input fields}

In the benchmark described in this paper, the dust grains are exposed to two sets of distinct radiation fields. The first set consists of eleven scaled versions of the Mathis ISRF, ranging from \emph{weak} to \emph{strong}, defined as
\begin{equation}
J^{\mathrm{SHG},i}(\lambda)=U_{i}\times J^{\mathrm{Mat}}(\lambda)\;\mathrm{with}\, U_{i}=10^{-4},10^{-3},...,10^{5},10^{6}
\end{equation}
The second set consists of the following six diluted black body fields with varying temperatures, ranging from \emph{soft} to \emph{hard}:
\begin{equation}
J^{\mathrm{SHG},j}(\lambda)=\begin{cases}
8.28\times10^{-12}\,B(\lambda,3000\,\mathrm{K}) \\
2.23\times10^{-13}\,B(\lambda,6000\,\mathrm{K}) \\
2.99\times10^{-14}\,B(\lambda,9000\,\mathrm{K}) \\
7.23\times10^{-15}\,B(\lambda,12000\,\mathrm{K}) \\
2.36\times10^{-15}\,B(\lambda,15000\,\mathrm{K}) \\
9.42\times10^{-16}\,B(\lambda,18000\,\mathrm{K}) \\
\end{cases}
\end{equation}
The dilution factors were chosen so that the far-infrared peak of the dust emissivity is at the same level for all fields in this set (for ease of visualization), and so that all fields in the set have a strength of $1\lessapprox U<10$.

\subsection{Calculation and wavelength grid\label{sec:wavegrid}}

Using the dust model described in section \ref{sec:model}, the codes participating in this benchmark calculate the spectral dust emissivity $\varepsilon(\lambda)$ for each of the input radiation fields specified in section \ref{sec:fields}, taking into account stochastic heating of small grains. The radiation emitted by the dust itself is ignored with respect to the input field, i.e.\ it is not the intention to calculate a self-consistent radiation field. The calculations are performed, and the results written down, using the wavelength grid on which the optical properties have been tabulated. This is a logarithmic grid with 1201 points in the range $0.001\,\micron\leq\lambda\leq10000\,\micron$.


\section{Dust emission\label{sec:emission}}

\subsection{Emission from a dust mixture\label{sec:mix-emission}}

The thermal emission of a dust grain depends nonlinearly on the grain size $a$, even in LTE conditions. It is therefore impossible to calculate the emission for a dust mixture with varying grain sizes from effective grain properties that would somehow represent the whole mixture \citep{2013ARAA..51...63S,2003ApJ...582..859W}. Instead, we define a grid of grain size bins $b$ for each dust model component $k$, and we choose an average, representative grain for each bin. We then proceed to calculate the emission as if each bin would contain only representative grains. For a sufficiently large number of bins, this procedure converges to the proper result.

A simple approach is to represent each bin by a grain size at the arithmetic or geometric center of the bin. In a somewhat more sophisticated approach, the absorption cross section per hydrogen atom representative for a particular bin can be calculated by an integration over the size distribution:
\begin{equation}
\varsigma_{k,b}^\mathrm{abs}(\lambda) = \int_{a^\mathrm{min}_{k,b}}^{a^\mathrm{max}_{k,b}}\pi a^{2}\, Q^{\mathrm{abs}}_k(a,\lambda)\, \Omega_k(a)\, \mathrm{d}a \label{eq:cross-section} \\
\end{equation}
where $[a^\mathrm{min}_{k,b},a^\mathrm{max}_{k,b}]$ specifies the size range of bin $b$ for dust model component $k$.

The representative mass of a dust grain in a particular bin can similarly be obtained from
\begin{equation}
M_{k,b}=  \left. \int_{a^\mathrm{min}_{k,b}}^{a^\mathrm{max}_{k,b}} \rho_k^{\mathrm{bulk}} \,\frac{4\pi}{3}\, a^{3} \,\Omega_k(a) \,\mathrm{d}a \;\; \middle/ \; \int_{a^\mathrm{min}_{k,b}}^{a^\mathrm{max}_{k,b}}\Omega_k(a)\,\mathrm{d}a \right.
\end{equation}
so that the enthalpy of a representative dust grain at temperature $T$ is given by 
\begin{equation}
 H_{k,b}(T)=M_{k,b}\, h_{k}(T)
\label{eq:grain-enthalpy}
\end{equation}
where $h_{k}(T)$ is the specific enthalpy of the grain material of dust component $k$ at temperature $T$.

The emissivity per hydrogen atom from a dust mixture with grain type components $k$ and grain size bins $b$ exposed to a radiation field $J(\lambda)$, called the \emph{input field}, can be expressed in function of the representative grain properties as 
\begin{equation}
\varepsilon(\lambda) = \sum_{k,b} \varsigma_{k,b}^\mathrm{abs}(\lambda) \int_0^\infty P_{k,b,J}(T) \,B(\lambda,T) \,\mathrm{d}T
\label{eq:grain-emission}
\end{equation}
where $B(\lambda,T)$ is the Planck function defined in Eq.\,\ref{eq:planck}, and $P_{k,b,J}(T)$ is the probability of finding the representative grain of bin $k,b$ at temperature $T$.

The emission originating from a dust mixture with specified total mass $M$ can then be written as 
\begin{equation}
j(\lambda)=\frac{M}{\mu}\,\varepsilon(\lambda)
\end{equation}
with $\mu$ given by Eq.\,\ref{eq:dustmass}. When combining the emission from various dust mixes, it is useful to recall that it is physically meaningful to add cross sections and masses, while mass coefficients, in general, cannot be added meaningfully:
\begin{equation}
\kappa_1+\kappa_2 =  \frac{\varsigma_1}{\mu_1} + \frac{\varsigma_2}{\mu_2} \ne  \frac{\varsigma_1+\varsigma_2}{\mu_1+\mu_2}.
\end{equation}

The challenge is thus to compute the probability distribution of grain temperatures, $P_{k,b,J}(T)$, which depends on the input radiation field in addition to the grain properties. See, for example, Fig. 4 of \citet{2007ApJ...657..810D} for an illustration of various temperature distribution curves.

In this discussion, we characterize $P$ as a function of grain temperature. The temperature of a grain and its internal energy are related through Eq.\,\ref{eq:enthalpy}, so we could equivalently characterize $P$ as a function of internal grain energy.

\subsection{Equilibrium heating dust emission\label{sec:equil-emission}}

When the representative grain in bin $k,b$ is in LTE with the surrounding radiation field $J(\lambda)$, the temperature probability distribution $P_{k,b,J}(T)$ becomes a delta function at the grain equilibrium temperature
\begin{equation}
P_{k,b,J}(T) = \delta(T - T^\mathrm{eq}_{k,b,J})
\end{equation}
and the equilibrium temperature can be determined via the energy balance equation
\begin{equation}
\int_{0}^{\infty} \varsigma^\mathrm{abs}_{k,b}(\lambda) \,J(\lambda) \,\mathrm{d}\lambda = \int_{0}^{\infty} \varsigma^\mathrm{abs}_{k,b}(\lambda) \,B(\lambda,T^
\mathrm{eq}_{k,b,J}) \,\mathrm{d}\lambda.
\label{eq:equilibrium}
\end{equation}

\subsection{Stochastic heating dust emission\label{sec:stoch-emission}}

When a single photon absorption may significantly change the internal energy of a representative grain, the grain is not in LTE with the surrounding radiation field $J(\lambda)$. The grain is \emph{stochastically heated}, and its state can no longer be characterized by a single temperature. In that case, we need to solve for the temperature probability distribution $P_{k,b,J}(T)$ to calculate the grain emission. The six RT codes benchmarked in this paper employ the method described in \citet{1989ApJ...345..230G,1992A&A...266..501S,1998A&A...337...85M} and \citet{2001ApJ...551..807D}. For ease of reference, this section summarizes the method using the quantities and notation introduced in the previous sections of this paper. We focus on a single grain size bin and a specific radiation field, dropping the indices $k,b$ and $J$ from the notation.

We select an appropriate temperature grid with $N$ bins $T_i$, $i=0,\dots,N-1$ (see Sect.\,\ref{sec:mix-emission}). Our goal is to determine the probabilities $P_i=P(T_i)$ for a grain to reside in temperature bin $i$.

We define a transition matrix $A_{f,i}$ describing the probability per unit time for a grain to transfer from initial temperature bin $i$ to final temperature bin $f$. The transition matrix elements in the case of heating $(f>i)$ are given by
\begin{equation}
A_{f,i}=4\pi\,\varsigma^{\mathrm{abs}}(\lambda_{fi})\, J(\lambda_{fi})\,\frac{hc\,\Delta H_{f}}{\left[H(T_f)-H(T_i)\right]^{3}}
\label{eq:coeff-heating}
\end{equation}
where $H(T_f)$ and $H(T_i)$ are the enthalpies of a dust grain in the final and initial temperature bins, $\Delta H_{f}=H(T_f^\mathrm{max})-H(T_f^\mathrm{min})$ is the enthalpy width of the final temperature bin, and $\lambda_{fi}$ is the transition wavelength which can be obtained from
\begin{equation}
\lambda_{fi}=\frac{hc}{H(T_{f})-H(T_{i})}.
\end{equation}
We assume that a dust grain cools by radiating photons with an energy that is very small compared to the internal energy of the grain. With this \emph{continuous cooling} approximation, cooling transitions occur only to the next lower level, so that $A_{f,i}=0$ for $f<i-1$ and
\begin{equation}
A{}_{i-1,i}=\frac{4\pi}{H(T_i)-H(T_{i-1})}\,\int_{0}^{\infty}\varsigma^{\mathrm{abs}}(\lambda)\, B(\lambda,T_{i})\,\mathrm{d}\lambda.
\label{eq:coeff-cooling}
\end{equation}
The diagonal matrix elements are defined as $A_{i,i}=-\sum_{f\ne i}A_{f,i}$, however there is no need to explicitly calculate these values as they are not used in the final procedure.

Assuming a steady state situation, the probabilities $P_{i}$ can be obtained from the transition matrix by solving the set of $N$ linear equations
\begin{equation}
\sum_{i=0}^{N-1}A_{f,i}\, P_{i}=0 \qquad f=0,...,N-1
\end{equation}
along with the normalization condition
\begin{equation}
\sum_{i=0}^{N-1}P_{i}=1
\end{equation}
where $N$ is the number of temperature bins. Because the matrix values for $f<i-1$ are zero these equations can be solved by a recursive procedure of computational order $\mathcal{O}(N^{2})$. To avoid numerical instabilities caused by the negative diagonal elements, the procedure employs a well-chosen linear combination of the original equations. This leads to the following recursion relations for the adjusted matrix elements $B_{f,i}$
\begin{align}
&B_{N-1,i} = A_{N-1,i} \qquad i=0,\ldots,N-2\\
&B_{f,i} = B_{f+1,i}+A_{f,i}, \quad f=N-2,\ldots,1;\, i=0,\ldots,f-1
\end{align}
and for the unnormalized probability distribution $X_{i}$
\begin{align}
X_{0} &= 1\\
X_{i} &=  \frac{\sum_{j=0}^{i-1}B_{i,j}X_{j}}{A_{i-1,i}}  \qquad i=1,\ldots,N-1
\end{align}
and finally for the normalized probabilities $P_{i}$
\begin{equation}
P_{i}=\frac{X_{i}}{\sum_{j=0}^{N-1}X_{j}} \qquad i=0,\ldots,N-1.
\end{equation}

While this method seems rather straightforward, specific algorithmic approaches differ between codes.

One important characteristic that tends to differ between implementations is the grid discretizing the grain temperatures (or equivalently internal energy states) during the construction of $P(T)$. With a fixed grid, a range of probable temperatures is defined, e.g. $2.7\,\mathrm{K} < T < T_\mathrm{max}$, where $T_\mathrm{max}$ is chosen to exceed the sublimation temperature of the bulk material being considered and the interval is divided in $N$ temperatures \citep[e.g.][]{2008AA...490..461B}. The distribution function is evaluated at that set of fixed internal energy states for all grains considered to be in the stochastic heating regime. The advantage of this approach is that many of the quantities used to generate $P(T)$ can be precomputed, reducing the computational requirements of the solution method. A disadvantage of this approach is that calculations are done over the full defined temperature range, including regions where $P(T)$ is negligible. Not only is this computationally inefficient, but it results in poor resolution of the form of $P(T)$, especially for grains of intermediate size where $P(T)$ will be relatively narrowly distributed with $T$. One alternate approach is to dynamically define the temperature grid \citep[e.g.][]{1999AA...349..907M,2001ApJ...551..277M}. In this iterative approach, a coarse and broad temperature grid is defined and $P(T)$ computed. The grid is refined based on $P(T)$; temperatures with low $P(T)$ are removed from the grid and $P(T)$ is recomputed on the new, smaller, more densely sampled grid. The grid refinement is continued until energy balance is achieved or the number of temperature points exceeds a predefined threshold. The advantage of this approach is that $P(T)$ is properly sampled for all grain sizes. The disadvantage is of course an increase in the computational load.  This is amplified by the fact that the algorithm will naturally increase the number of temperature samples as the grain approaches the equilibrium regime, because $P(T)$ is increasingly peaked as the grain size increases. 

A second important characteristic that tends to differ between implementations is the mechanism to transition from stochastic to equilibrium heating regime. The simplest approach is to fix the grain size of the transition so that all grains with $a < a_\mathrm{trans}$ are considered to be stochastically heated and all those with $a \ge a_\mathrm{trans}$ are considered to be in the equilibrium regime, regardless of the true state of the grain. Since the appropriate transition point is a function of the radiation environment in addition to composition, this approach leads to errors in the treatment of the emission. However, with judicious selection of $a_\mathrm{trans}$ , e.g. $a_\mathrm{trans} \sim 0.01$\,\micron \, \citep{2008AA...490..461B}, the results can be acceptable at least for non-extreme field strengths. Alternatively, the characteristics of $P(T)$ can be used to terminate the stochastic heating treatment and transition to equilibrium heating. For example, in the case of a dynamic temperature grid as described above, the size at which the heating algorithm fails can be defined as $a_\mathrm{trans}$ \citep{2001ApJ...551..277M}. A third method to determine $a_\mathrm{trans}$ is to compute the absorption and radiative timescales for each grain size in the considered radiation field \citep{2001ApJ...551..807D}. These timescales are a natural physical metric as stochastic heating occurs when the mean time between photon absorptions is long compared to the time the grain takes to radiatively cool. The ensemble of grains will then be found at a large range of temperatures resulting in a broad probability distribution. With this approach, if the absorption timescale is significantly shorter than the radiative timescale at a given grain size, the stochastic heating regime is terminated for that and all larger sizes.

A third characteristic that may differ between implementations is the discretization of the grain size distribution, as discussed in Sect.\,\ref{sec:mix-emission}.


\begin{table*}
\caption{Overview of the discretization parameters and heuristics used by participating codes. Refer to Sects.\,\ref{sec:code-skirt} through \ref{sec:code-dartray} for a more extensive description. Also, Fig.\,\ref{fig:a-trans} further supplements the information in the last column of this table.}
\label{tbl:heuristics}
{\renewcommand{\arraystretch}{1.15}
\begin{tabular}{lccll}
\hline\hline
Code & Grain size bins & Temperature & Heuristic to select or determine    & Heuristic to transition from  \\
         & (Sil/Gra/PAH)      & bins               & temperature grid       & stochastic to equilibrium regime \\
\hline
SKIRT      & 15/15/15    &  20/625/1250 & one of 3 grids based on width of $P(T)$    & based on width of $P(T)$ \\
DIRTY      & 121/121/28 &  50-1000      & iterative range \& resolution adjustment     & based on $\tau_{\rm abs}$/$\tau_{\rm rad}$ \\
TRADING & 20/20/8      &  80               & fixed predefined grid                                & fixed at $a_\mathrm{trans}=0.05\,\micron$ \\
CRT         & 15/15/15    &  128          & one of 6 grids based on $\mathcal{P}_\mathrm{abs}$  & based on $\mathcal{P}_\mathrm{abs}$ \\
MCFOST  & 63/62/24    &  300             & fixed predefined grid                                & based on $\tau_{\rm abs}$/$\tau_{\rm rad}$ \\
DART-Ray & 63/62/24   &  200             & iterative range adjustment                        & based on $\sigma_T$ of Gaussian approx. \\
\hline
\end{tabular}}
\end{table*}

\section{Reference code and participating codes\label{sec:codes}}

\subsection{DustEM\label{sec:code-dustem}}

The DustEM code is described in \citet{2011AA...525A.103C}. DustEM is a stand-alone code (i.e.\ it is not a RT code) that calculates the emission and extinction of dust grains given their size distribution and their optical and thermal properties. It determines the grain temperature distribution $P(T)$ using the formalism of \citet{1986AA...160..295D}, and it then computes the dust SED and associated extinction for given dust types and size distributions. To correctly describe the dust emission at long wavelengths the original algorithm has been adapted to better cover the low temperature region. Using an adaptive temperature grid, DustEM iteratively solves the integral equation Eq. 25 from \citet{1986AA...160..295D} in the approximation where the grain cooling is fully continuous. The temperature distribution calculation is performed for all grain populations and sizes including those for which the thermal equilibrium approximation would apply.

We produced the reference solutions in this paper with the public version of DustEM (v3.8, dated Spring 2010). To this end, we converted the dust properties defined in Sect.\,\ref{sec:model} and the input radiation fields defined in Sect.\,\ref{sec:fields} into the data format expected by DustEM. We adjusted the values of the physical constants in the DustEM code, raised the maximum number of grain size and temperature bins to accommodate our input data, and fixed a minor issue in the routine that imports the grain size distribution. More importantly, to obtain accurate reference solutions, we substantially increased the number of temperature bins and the number of numerical iterations (see Sect.\,\ref{sec:reference-solutions}). Other than this, the DustEM code was used without modifications.

Our use of DustEM for producing reference solutions should not be understood to imply that it necessarily produces the physically correct results. The DustEM implementation relies on the continuous cooling assumption just like the RT codes participating in this benchmark. However, since it is not a RT code by itself, DustEM's focus is solely on calculating dust properties and emission, and it has a neutral status in the context of this benchmark.

The following sections describe each of the participating codes, with a focus on the specific heuristics employed for calculating the results presented in this paper. Table \ref{tbl:heuristics} offers a (very) concise overview of this information.

\subsection{SKIRT\label{sec:code-skirt}}

SKIRT \citep{2003MNRAS.343.1081B,2011ApJS..196...22B,CampsBaes2015} is a Monte Carlo continuum RT code for simulating the effect of dust on radiation in astrophysical systems. It offers full treatment of absorption and multiple an\-isotropic scattering by the dust, computes the temperature distribution of the dust and the thermal dust re-emission self-con\-sistently, and supports stochastic heating of small grains using an efficient library approach. The code handles multiple dust mixtures and arbitrary 3D geometries for radiation sources and dust populations, and offers a variety of simulated instruments for measuring the radiation field from any angle. It features a wide range of built-in components that can be configured to construct complex models without changing or adding source code.

SKIRT closely follows the method presented in Sect.\,\ref{sec:emission} to calculate dust emission. The computation time for the emissivity of a stochastically heated dust mixture exposed to a particular radiation field scales roughly with $B N^2$, where $B$ is the total number of size bins in the dust mixture (for all grain types combined), and $N$ is the number of temperature bins used in the calculation.

For the results shown in this paper, we use 15 size bins $b$ for each grain type $k$, distributed logarithmically over the complete size range, so that $B=45$. The absorption efficiencies loaded from the tables described in Sect.\,\ref{sec:model-data} are interpolated logarithmically as needed to perform the integrations over grain size presented in Sect.\,\ref{sec:mix-emission} over a logarithmic grain size grid with 201 points within each bin.

Because of the $N^2$ dependency in the computation time, the choice of the temperature grid is rather crucial. By varying $N$ in a number of experiments, it can be easily shown that, for most input fields, the temperature probability distribution $P(T)$ for very small grains can be calculated accurately on a rather coarse grid. Larger grains require a finer grid because $P(T)$ is narrower and has steep flanks. As discussed in Sect.\,\ref{sec:equil-emission}, for sufficiently large grains $P(T)$ approaches a delta function, so that the procedure described in Sect.\,\ref{sec:stoch-emission}  requires an exceedingly refined grid with $N>5000$ to produce accurate results, which becomes computationally prohibitive. Thus, in the interest of both speed and accuracy, we need to switch from transient to equilibrium calculations for grains that can be considered to be in LTE.

A RT simulation typically calculates the emissivity of a certain (fixed) dust mixture for a large number of radiation fields. This computation can be accelerated substantially by pre-calculating and storing the elements of the matrix $A_{f,i}$ defined in Sect.\,\ref{sec:stoch-emission}, insofar as they don't depend on the radiation field. The memory requirements scale with $B N^2$, just like the computation time. More importantly, pre-calculating these values requires a predefined temperature grid that remains fixed for all emissivity calculations. However, performing all calculations on the finest grid would be very inefficient. 

The SKIRT implementation handles these conflicting requirements as follows. We predefine three separate temperature grids (A, B, and C) that can be used for any of the emissivity calculations, and we pre-calculate and store all radiation-field-independent values on each of these three grids for each size bin. The temperature range is the same for all grids; it is usually set to $[2\,\mathrm{K},3000\,\mathrm{K}]$ but if needed the upper limit is decreased to the largest temperature for which enthalpy data is available in the dust properties.

Grid A has only 20 bins. The widths are distributed according to a power law, providing a lot more resolution at low temperatures where most of the action is. The ratio of the largest bin (at high temperatures) over the smallest bin (at low temperatures) is set to 500. This grid is used to find a quick estimate of the range in which the temperature probability distribution is nonzero (or rather, larger than a very small fraction of the maximum probability).

All bins in grid B are 4\,K wide. This medium-resolution grid is used to calculate the temperature probability distribution for dust grains with a very wide temperature range (i.e. very small grains and essentially all PAH grains).

Grid C has an average bin width of 2\,K, with a power-law ratio of 3 between the largest and smallest bins. This provides a fine resolution of less than 1\,K at low temperatures while still offering decent resolution at high temperatures. This high-resolution grid is used to calculate the temperature probability distribution for dust grains with a rather narrow temperature range (but not so narrow that they would be considered to be in equilibrium).

SKIRT implements the following heuristic to select the appropriate calculation for each representative dust grain:
\begin{enumerate}
\item calculate the equilibrium temperature $T_{\mathrm{eq}}$ for this grain;
\item use grid A to estimate the temperature range $\Delta T=T_{\mathrm{max}}-T_{\mathrm{min}}$ in which $P(T)/P_\mathrm{max} > 10^{-20}$;
\item \label{item:testEQ1} if $\Delta T<10\,\mathrm{K}$ or if $T_{\mathrm{max}}<T_{\mathrm{eq}}$ then calculate the emissivity assuming equilibrium at temperature $T_{\mathrm{eq}}$ and exit;
\item \label{item:finalPT} calculate $P(T)$ using grid B (if $\Delta T>200\,\mathrm{K}$) or grid C (if $\Delta T<200\,\mathrm{K}$);
\item \label{item:finalDT} update the temperature range $\Delta T=T_{\mathrm{max}}-T_{\mathrm{min}}$ in which $P(T)/P_\mathrm{max} > 10^{-20}$ based on the new calculation;
\item \label{item:testEQ2} if $\Delta T<10\,\mathrm{K}$ or if $T_{\mathrm{max}}<T_{\mathrm{eq}}$ then calculate the emissivity assuming equilibrium at temperature $T_{\mathrm{eq}}$ and exit;
\item calculate the emissivity using $P(T)$ from step \ref{item:finalPT} over the range $[T_{\mathrm{min}},T_{\mathrm{max}}]$ determined in step \ref{item:finalDT}.
\end{enumerate}
The conditions in steps \ref{item:testEQ1} and \ref{item:testEQ2} are designed to avoid numerical instabilities when the temperature probability distribution approaches a delta function ($\Delta T<10\,\mathrm{K}$), and to capture situations where the result is clearly inaccurate since the equilibrium temperature lies outside the calculated temperature range ($T_{\mathrm{max}}<T_{\mathrm{eq}}$).

Further experiments with the SKIRT implementation show that for $\Delta T\gtrsim25\,\mathrm{K}$ the result is highly sensitive to the exact value of $\Delta T$; for smaller values the result converges to a stable solution. For values down to $\Delta T\thickapprox10\,\mathrm{K}$, the result is numerically stable in the sense that performing the calculation on higher-resolution grids produces essentially the same solution.

\subsection{DIRTY\label{sec:code-dirty}}

DIRTY \citep{2001ApJ...551..269G, 2001ApJ...551..277M} is a Monte Carlo RT code designed to study dust and its effect on radiation in arbitrary astrophysical systems. DIRTY is a fully 3D code allowing for the specification of arbitrary density distributions of both dust and radiation sources. It implements an adaptive mesh allowing for the efficient allocation of computing resources amongst regions in the model space depending on the physical characteristics of the system. Dust absorption, temperature distribution, and emission are handled self-consistently and multiple, anisotropic scattering is implemented. The dust heating implementation supports both equilibrium and stochastic processes based on the local radiation field and dust properties at each grid in the model space. 

Like other codes presented here, DIRTY follows the approach presented in Sect.\,\ref{sec:emission}. Internal to DIRTY, the dust grain size distribution is not further discretized beyond the input discretization of the model; for the benchmark dust model, we compute the heating and emission for all sizes in the input mesh (28 for PAH, 121 for graphitic and silicate components; see Sect.\,\ref{sec:model-data}).  

The heuristics employed by DIRTY in calculating the dust emission from each grain size of each component exposed to the local radiation field at a point in the model space are as follows: 

\begin{enumerate}
\item \label{heur:timeScales} The equilibrium temperature, $T_{\rm eq}$, cooling timescale, $\tau_{\rm rad}$, and heating 
  time scale, $\tau_{\rm abs}$ are computed according to section 7 of \citet{2001ApJ...551..807D};
\item \label{heur:stochasticTerm} If the time scales computed in step \ref{heur:timeScales}
  satisfy the inequality $\tau_{\rm abs} < \tau_{\rm rad}$, the grain is considered to be in
  equilibrium with the local field; it is assigned a temperature of $T_{\rm eq}$ and its emission
  is calculated following Eq. \ref{eq:grain-emission} with $P=\delta(T-T_{\rm eq})$.  All grains of the same
  composition larger than the size for which the inequality is first satisfied are by
  default treated as being in equilibrium with the local field. 
\item \label{heur:coarseGrid} For those grains found to be in stochastic regime in step
  \ref{heur:stochasticTerm}, $P(T)$ is computed on an initial coarse temperature grid. The
  coarse grid is defined on 50 points linearly spaced between $\left[ 0.3,3.0 \right] T_{\rm eq}$. 
\item \label{heur:coarseGridRefine} Depending on the results of step
  \ref{heur:coarseGrid}, the algorithm proceeds in one of two directions:
  \begin{enumerate}
  \item \label{heur:coarsePtolHigh} If $P(T)$ is not below a specified tolerance at the
    endpoints of the temperature grid ($P(T_{\rm min}),P(T_{\rm max}) < 10^{-15}$), the
    temperature limits on the grid are
    expanded by 50\% and we return to step \ref{heur:coarseGrid}
    with the new, expanded temperature grid.
  \item \label{heur:coarsePtolGood} If $P(T)$ is below the tolerance at the endpoints of
    the temperature grid, compute the energy emitted by the stochastically
    heated grain using Eq. \ref{eq:grain-emission}.  If the emitted energy is within 1\% of the energy
    absorbed from the radiation field by the grain, consider the calculation converged
    and return the emitted energy spectrum. Otherwise, proceed to step \ref{heur:gridRefine}. 
  \end{enumerate} 
\item \label{heur:gridRefine} The grid is now refined by a series of moves that refine the
  temperature limits and increase the number of temperature samples if necessary.  
  \begin{enumerate} 
  \item \label{heur:gridRefineTrim} Remove all points on the ends of the temperature range
    which have small probabilities ($P(T) < 10^{-15}$), 
  \item \label{heur:gridRefineProb} Recompute the probabilities on the smaller grid with
    the same number of samples.  
  \item \label{heur:gridRefineEmit} Compute the emission and emitted energy. If the
    emitted energy matches the absorbed energy to 1\%, consider the calculation converged
    and return the emitted energy spectrum. Otherwise; 
  \item \label{heur:gridRefineIncrease} Keeping the same temperature limits, increase the
    number of samples by 50\% and recompute the probabilities. Trim temperature endpoints
    with $P(T) < 10^{-15}$ and return to step \ref{heur:gridRefineProb}.  If this step
    would result 
    in the number of bins exceeding a pre-defined maximum (1000), record the failure of
    the stochastic heating algorithm and proceed to the next grid in the model space.
  \end{enumerate}
\end{enumerate}

In practice, the failure described in step \ref{heur:gridRefineIncrease} occurs in model bins for which the local field has not been well defined, generally due to very few photons interacting in that cell, either through a poor definition of the adaptive mesh or insufficient photons being run in the Monte Carlo simulation of the radiation. These cells are generally unimportant in the overall energy budget of the model and can be masked in post processing. Such cells are rare in most model runs; their number and distribution in the model space can be used as a metric of the overall quality of the simulation. 

The approach in step \ref{heur:stochasticTerm} results in all PAH sizes being treated in the stochastic regime in the radiation fields explored in this benchmark. Silicate and graphitic grains generally achieve equilibrium between sizes of 0.006 and 0.020\,\micron.

\subsection{TRADING\label{sec:code-trading}}

TRADING \citep{2008AA...490..461B} is a 3D Monte Carlo dust continuum RT code with characteristics similar to those of SKIRT and DIRTY. Originally designed to study the effects of clumping in the disks of spiral galaxies, it uses an binary-tree adaptive grid (octree) for the dust distribution.

TRADING computes stochastic heating following the method described in Sect.\,\ref{sec:emission}. A single temperature grid is used to precompute
the field-independent terms of the matrix elements (Eq.\,\ref{eq:coeff-cooling} and most factors in Eq.\,\ref{eq:coeff-heating}). Since the RT models of clumpy galactic disks in \citet{2008AA...490..461B} need a few million dust cells, and each cell requires the calculation of dust heating for the full grain size distribution, \citet{2008AA...490..461B} uses a limited temperature grid of 80 logarithmically spaced bins between 2.7 and 2000\,K. As mentioned in Sect.\,\ref{sec:stoch-emission}, faster thermal equilibrium calculations were performed for grains larger than $a_\mathrm{trans}=0.01$\,\micron. The original setup resulted in SEDs that were estimated to be within 10\% of full solutions for wavelengths up to 1000\,\micron.

For this benchmark we left the number of temperature bins at 80, but we extended the temperature range to 3000\,K. While the previous choice of $a_\mathrm{trans}$ produces accurate results for the typical interstellar radiation fields encountered in spiral galaxies (with $U\gtrsim 0.1$, see \citet{2012ApJ...756..138A,2014arXiv1409.5916H}), we adopted here $a_\mathrm{trans}=0.05$\,\micron\ in order to improve the solution for fields with lower intensities.

The dust grain size distribution was discretized using a grid with 20 size bins for graphite, 20 for silicates, and 8 for PAHs, logarithmically spaced over their size range. The choice allows to have similar bin widths for all materials. Optical properties were derived interpolating logarithmically over those of the full size table. The grain size grid used for this paper is similar to that adopted in \citet{2008AA...490..461B} though for a different dust model.

For $a_\mathrm{trans}=0.01$\,\micron, only about half of the size bins pass through the stochastic heating calculation. For $a_\mathrm{trans}=0.05$\,\micron, this goes up to 75\%: all of the PAH bins and 14 of the 20 graphite and silicate bins. If a grain with $a<a_\mathrm{trans}$ attains the condition of thermal equilibrium, the adopted temperature grid might fail in computing the resulting narrow probability function. Thus, in addition to the grain size cut-off, we chose to assume thermal equilibrium whenever the range of the computed temperature distribution defined by $P(T) > 10^{-15}$ is smaller than the equilibrium temperature.

\subsection{CRT\label{sec:code-crt}}

CRT \citep{2003AA...397..201J, 2005AA...440..531J, 2012AA...544A..52L} is a Monte Carlo dust continuum RT program. It solves the RT equation self-consistently with a full treatment of scattering, absorption and emission of radiation in 1D, 2D, and 3D geometries. The program allows using an external component, for instance DustEM, for the calculation of the dust emission spectrum for a given input radiation field. This feature has been used, e.g.\ in \citet{2011AA...535A..89Y}, to study the microwave emission from spinning dust grains. However, CRT itself also has optimized routines for fast dust emission calculations, including the treatment of stochastically heated grains. Although CRT can calculate dust emission with fully discrete cooling, the results presented in this paper use the continuous cooling approximation, and the following discussion is focused on the algorithms for continuous cooling computations.

The basic algorithms employed by CRT follow the outline presented in Sect.\,\ref{sec:emission}. To allow the use of pre-computation to speed up the construction of the transition matrix $A$, the temperature grid is not defined fully dynamically according to the input field. Instead, each dust type and size uses one of several predefined grids, for which the pre-computations are done at the beginning of calculation. The predefined grids are linear in $T$ and their upper and lower limits are chosen to allow a good representation of the grain temperature distribution in the types of radiation fields that are found in the model. In particular, the grids should be built for reference fields that span approximately the range of radiation energy densities expected to be found in the model. If the hardness of the radiation field varies significantly, it may also be useful to include grids for different spectra with the same energy density. For the calculations presented in this paper we use six temperature grids that were built for scaled Mathis ISRF with $U=10^{-4}, 10^{-1}, 1, 10, 10^3$, and $10^6$. For large grains and strong radiation fields, the predefined grid has a special entry that triggers equilibrium calculation, otherwise full stochastic calculations is used. The number of temperature bins for stochastic calculation can be set by the user; in this paper we use 128 bins.

To select the temperature grid that is used for calculating the emission for a grain size and type in a given input field, we calculate the absorption time scale $\tau_{\mathrm{abs}}$ and the mean energy of the absorbed photon $\left<E_{\mathrm{abs}}\right>$. We choose the predefined grid that has been built for the reference field most like the current input field. In the calculations presented in this paper the selected grid is simply the one for whose reference field the mean absorbed power per grain $\mathcal{P}_{\mathrm{abs}}=\left<E_{\mathrm{abs}}\right>/\,\tau_{\mathrm{abs}}$ is closest to the values calculated for the input field. Although the selected temperature grid is not necessarily optimal, it is good enough to allow accurate results using a modest number of energy bins.

The computation of emission from stochastically heated grains in CRT differs slightly from the description given in Sect.\,\ref{sec:stoch-emission}. Instead of using Eq.\,\ref{eq:coeff-heating} for calculating the upwards transition rates, we apply Eqs.\,15--25 and 28 from \citet{2001ApJ...551..807D}, which include corrections for the finite size of an enthalpy bin. Similarly, instead of Eq.\,\ref{eq:coeff-cooling}, we use Eq.\,41 from \citet{2001ApJ...551..807D} for calculating the cooling part of the transition matrix. Including these finite bin size corrections allows using a lower resolution grid, which substantially benefits the computation time.

The cooling matrix elements $(f<i)$ are independent of the radiation field and they are pre-computed. The heating elements $(f>i)$ depend on the radiation field and must be calculated separately for each input field. Moreover, when using the equations from \citet{2001ApJ...551..807D}, instead of using the radiation field strength at a single wavelength as in Eq.\,\ref{eq:coeff-heating}, we must integrate numerically over the wavelength grid. We use precomputed integration weights $w_{f,i}$ corresponding to each grid point of the wavelength grid. The integral in Eq.\,15 in \citet{2001ApJ...551..807D} can then be evaluated as $A_{f,i}=\sum_{k=1}^{N_\lambda} w_{f,i}(k)J(\lambda_k)$. If the number of points in the wavelength grid is large, calculating the full sum is slow. However, for given energy bins $i$ and $f$, radiation within only a narrow range of wavelengths can induce transitions $i\rightarrow f$. Therefore, $w_{f,i}(k)$ is non-zero only for a few $k$ and only non-zero integration weights are stored and used in the summation.

Discretization of the particle size distribution can be defined by the user. In this paper we employ the same discretization as SKIRT, i.e., 15 logarithmic bins for each of the three dust components. Dust properties for each size bin are calculated according to Eqs.\,\ref{eq:cross-section}-\ref{eq:grain-enthalpy} using numerical integration with 256 grid points.

\subsection{MCFOST\label{sec:code-mcfost}}

MCFOST \citep{2006AA...459..797P,2009AA...498..967P} is a 3D continuum and line RT code. It relies on the Monte Carlo method to compute the local specific intensities and related quantities (e.g. temperature, molecular levels) and computes observables via a ray-tracing method. The emerging fluxes are calculated by formally integrating the RT method along rays using the specific intensities and source functions computed during the Monte Carlo run.

MCFOST computes the stochastic heating of small dust grains following the method presented in section\,\ref{sec:emission}, with a few refinements to ensure numerical stability and speed.

We first compute the time between two successive absorptions of a photon and compare it to the cooling time of the grain, following the method described in \citet{2001ApJ...551..807D}. For dust grains where the time between two absorptions is smaller than the cooling time, only the equilibrium temperature is calculated.

For those grains which have a shorter cooling time, we compute the full temperature probability distribution using a fixed temperature grid with 300 points logarithmically distributed between 1 and 3000\,K. The cooling terms of the transition matrix (Eq.\,\ref{eq:coeff-cooling}), which are independent of the radiation field, are precomputed, as well as most of the heating terms factors (Eq.\,\ref{eq:coeff-heating}). We estimate the specific intensity $J(\lambda_{fi})$ for each term in Eq.\,\ref{eq:coeff-heating} by interpolating the radiation field computed by the Monte Carlo run. As the interpolation coefficients are identical for every cell in the model, they are also precomputed to speed-up the calculations.

For dust in radiative equilibrium, MCFOST solves the problem of self-consistent dust heating and re-emission using the immediate re-emission concept of \citet{2001ApJ...554..615B}. This methods eliminates the need for iteration and ensures a perfect conservation of energy. For non-equilibrium dust grains however, this procedure is prohibitive as it requires a temperature calculation at each absorption/re-emission. Instead, we use the classical iterative scheme where we store the energy absorbed by the dust, compute the temperature probability distribution in all cells once all the packets have been propagated and re-emit the absorbed energy via new packets according to the new temperature probability. The procedure is iterated until a desired convergence on the temperature or energy is reached. Because only the fraction of radiation that is absorbed by the non-equilibrium grains needs to be re-emitted in an iterative way, convergence is usually reached after only a few iterations. In practice, when a packet is absorbed inside a cell, its energy is split: the fraction absorbed by dust grains in equilibrium is immediately re-emitted, while the fraction absorbed by non-equilibrium grains is stored to be re-emitted during the next iteration. For the benchmark presented in this paper, the input radiation field is fixed so no iteration is required.

In the presented calculations, the grain size distributions were discretized using 63 logarithmically spaced grain sizes for silicates, 62 for graphite and 24 for PAHs.

\subsection{DART-Ray\label{sec:code-dartray}}

DART-Ray is a ray-tracing 3D dust RT code that implements the RT algorithm described in \citet{2014MNRAS.438.3137N}. It can be used to derive radiation field energy density distributions and outgoing radiation surface brightness maps for arbitrary 3D distributions of dust mass and stellar emission. It includes treatment of both absorption and anisotropic scattering. For the dust emission calculations, DART-Ray uses the prescription initially incorporated in the 2D RT model of \citet{2000A&A...362..138P}, and later updated in \citet{2011A&A...527A.109P}. However, unlike the 2D models where the stochastically heated dust emission could be explicitly computed for each individual position, the calculations for the 3D models, containing of the order of $10^6$ cells, can be accelerated by using an adaptive SED library approach (see Natale et al. 2014b, in press).

For a given radiation field intensity spectrum, found for a particular model cell, the stochastically heated dust emission is derived following \citet{1991ApJ...379..122V}. The method used to determine the probability distribution $P(T)$ combines the numerical integration of \citet{1989ApJ...345..230G} with a step-wise analytical solution. The algorithm provides accurate and swift results on a relatively coarse grid. This is particularly useful for larger grain sizes where the probability distribution $P(T)$ converges to a narrow distribution around the equilibrium temperature. In case of the first order integration of \citet{1989ApJ...345..230G}, the width of the energy bins needs to be considerably smaller than the mean deposited energy to preserve energy balance. An increasing number of energy bins would be required to avoid energy losses in the heating process, which would make the calculation inefficient \citep[see][and Sect.\,\ref{sec:stoch-emission}]{FischeraPhD}.

The absorption of CMB photons is assumed to provide a continuous heating source. It is taken into consideration by subtracting the heating rate related to the CMB from the cooling rate (Eq.\,\ref{eq:coeff-cooling}), which limits the temperatures to values not lower than the CMB temperature. As further discussed in Sects.\,\ref{sec:benchmark-solutions} and \ref{sec:weak-fields}, the cooling assumed in the dust model is only valid as long as the emitted photon energy of the modified black body is small relative to the enthalpy of the grain. This assumption does no longer apply at very low temperatures of small dust grains or PAH molecules \citep{FischeraPhD}. While negligible for most cases we find that the CMB becomes a considerable heating source for the very low radiation field strength. For $U=10^{-4}$ the heating of silicate grains by CMB photons is for all sizes larger than 10\%. For graphites the contribution is considerably lower with only a few percent and still negligible for PAH molecules.

The temperature distribution $P(T)$ for each grain size of a given composition is obtained consecutively by starting with the largest grain size and utilizing the basic characteristic that the distributions broaden with decreasing grain size. The probability distribution is only considered for all grains below a certain size where the stochastic heating leads to a considerable distribution of the dust temperatures. Above this critical grain size the grains are assumed to radiate at the equilibrium temperature $T_{\rm eq}$. To estimate the transition from equilibrium to non equilibrium, we apply the Gaussian approximation in the limit of large grains as derived by \citet{1991ApJ...379..122V}. We consider the grains to be stochastically heated if $2\sigma_T/T_{\rm eq} > 0.05$. For $0.05 < 2\sigma_T/T_{\rm eq}<0.1$ we apply the Gaussian approximation; for $2\sigma_T/T_{\rm eq}>0.1$ the full $P(T)$ distributions are derived.

The temperature distributions of stochastically heated dust grains are derived for dynamically determined temperature intervals [$T_{\rm min}$, $T_{\rm max}$]. For the results presented in this paper, we subdivide the temperature interval into 200 temperature bins equally spaced in $\log T$. The interval boundaries are determined iteratively by increasing $T_{\rm min}$ or decreasing $T_{\rm max}$ by $30\%$ until the probabilities $P(T)$ for the lowest and highest temperature bin are lower than $10^{-20}$, ensuring that the emission at higher and lower temperatures can be neglected. To accelerate the iterative process, we use for each dust grain as the initial estimate of the temperature interval the derived interval of the previous larger dust grain. For the first grain for which the temperature distribution is derived, we use as initial estimate of the temperature interval a width based on the Gaussian approximation in the limit of large dust grains.

For the results presented in this paper, we derive the dust emission for each grain species at each grain size of the tabulated size distribution described in Sect.\,\ref{sec:model-data}. To calculate the total emission, we then integrate over the size distribution and sum the contributions from different grain species. By comparing the total dust emission with the total absorbed energy we ascertain that the energy balance for every grain species is fulfilled with an accuracy better than a few percent. The largest discrepancies (3-4\%) are found for the most extremely scaled Mathis fields considered in the benchmark ($U=10^{-4}$ and $U=10^6$).


\section{Results and discussion\label{sec:results}}

\subsection {Data files\label{sec:result-data}}

The data files representing the benchmark results can be downloaded from the web site indicated in footnote \ref{fn-website}. For each participating code, and for each input radiation field, the calculated solution is stored in a separate text file with columns specifying the wavelength $\lambda$ (in \micron); the mean intensity $J_\lambda$ of the input field (in $\mathrm{W}\,\mathrm{m}^{-3}\,\mathrm{sr}^{-1}$); and the silicate, graphite and PAH emissivities $\lambda\,\varepsilon_\lambda$ (in $\mathrm{W}\,\mathrm{sr}^{-1}\,\mathrm{H}^{-1}$), in that order. The file naming scheme and the precise file format are described on the web site.

\subsection{Reference solutions\label{sec:reference-solutions}}

As mentioned in Sect.\,\ref{sec:code-dustem}, below we compare the results from the codes participating in this benchmark against reference solutions generated with DustEM. Figure \ref{fig:dustem-solutions} shows these solutions for a selection of the input fields defined in Sect.\,\ref{sec:fields}.

To ensure proper accuracy, we increased the number of temperature bins and the number of iterations in the DustEM integral equation solver until the calculated emission converged to a stable solution; see Fig.\,\ref{fig:dustem-convergence}. Specifically, the number of temperature bins was raised from 200 to 3500, and the number of iterations from 80 to 250. These changes dramatically increase the computation time, however this is acceptable for the calculation of a reference solution.

As an extra sanity check, we verified that the emissivities calculated by DustEM (using the same number of temperature bins and iterations as for the reference solutions) indeed converge to the corresponding equilibrium emissivities. Fig.\,\ref{fig:dustem-equilibrium} shows this comparison for 0.05\,\micron\ grains exposed to radiation fields ranging from extremely weak (left) to strong (right). For a strong field, where we expect the grains to be in equilibrium, the DustEM solutions indeed match the equilibrium emissivities. For a weak field, the solutions differ since the grains are no longer completely in equilibrium. This shows the importance of performing the full stochastic calculation in the presence of extremely weak fields, even for grain sizes up to 0.05\,\micron. Codes transitioning from stochastic to equilibrium calculation at a fixed grain size should thus set a sufficiently high value of $a_\mathrm{trans}$ (see Sects.\,\ref{sec:stoch-emission} and \ref{sec:code-trading}).

\subsection{Benchmark solutions\label{sec:benchmark-solutions}}

Fig.\,\ref{fig:benchmark-total} compares the total emissivities calculated by each of the codes participating in this benchmark to the corresponding reference solutions for a selection of the input fields defined in Sect.\,\ref{sec:fields}. Subsequent figures zoom in on the emissivities for each dust component separately, i.e.\ silicate (Fig.\,\ref{fig:benchmark-silicate}), graphite (Fig.\,\ref{fig:benchmark-graphite}), and PAH (Fig.\,\ref{fig:benchmark-pah}) grains.

In these figures we limit the displayed wavelength range to $1\,\micron\le\lambda\le1000\,\micron$. Outside of this range, other sources or processes usually dominate the radiation emanating from astrophysical objects, so it is not relevant to evaluate the results of a dust emissivity calculation in that spectral range. Moreover, some of the assumptions underlying the computations are no longer valid, rendering the results physically meaningless.

First considering the shorter wavelength range, a black body with peak emission at $\lambda=1\,\micron$ has a temperature of $T\approx2900\,\mathrm{K}$. The sublimation temperature of a dust grain is estimated at $1200\,\mathrm{K}$ for silicates and at $2100\,\mathrm{K}$ for graphites \citep{2009Icar..201..395K}. Evaporation rates rise roughly exponentially with increasing grain temperature \citep{1989ApJ...345..230G,2009Icar..201..395K}, i.e.\ with decreasing emission wavelength. It is clear that a relevant fraction of the dust that would emit at $\lambda\lesssim1\,\micron$ is destroyed by evaporation, so that the grain size distribution in our dust model is no longer valid under these conditions. Consequently, the calculation would substantially overestimate the resulting dust emission.

We now consider the longer wavelength range. For a sufficiently strong input field, say $U\gtrsim1$, we can expect the calculated emissivity results to be correct because most of the grains emitting at wavelengths longer than $1000\,\micron$ are in LTE. However, the emissivity peaks at much shorter wavelengths, so that the level at $1000\,\micron$ is already several orders of magnitude below the peak level (see all panels in Fig.\,\ref{fig:dustem-solutions} except for the first two). The situation is different for extremely weak input fields approaching the level of the cosmic microwave background (CMB). A black body with peak emission at $\lambda=1000\,\micron$ has a temperature of $T\approx2.9\,\mathrm{K}$, just above the temperature of the CMB. For small dust grains at such low energies, the continuous cooling assumption no longer holds,\footnote{With the properties of our dust model, the internal energy of a 0.007\,\micron\ graphite grain at 2.9\,K is insufficient to emit a $1000\,\micron$ photon.} meaning that the method presented in Sect.\,\ref{sec:stoch-emission} does not necessarily yield correct results. In conclusion, the dust emission at wavelengths $\lambda\gtrsim1000\,\micron$ is either calculated assuming equilibrium conditions (which renders comparison uninteresting) or calculated improperly (which renders comparison meaningless).

\subsection{Evaluation of benchmark results\label{sec:eval-results}}

\begin{figure*}[p]
  \centering
  \includegraphics[width=\textwidth]{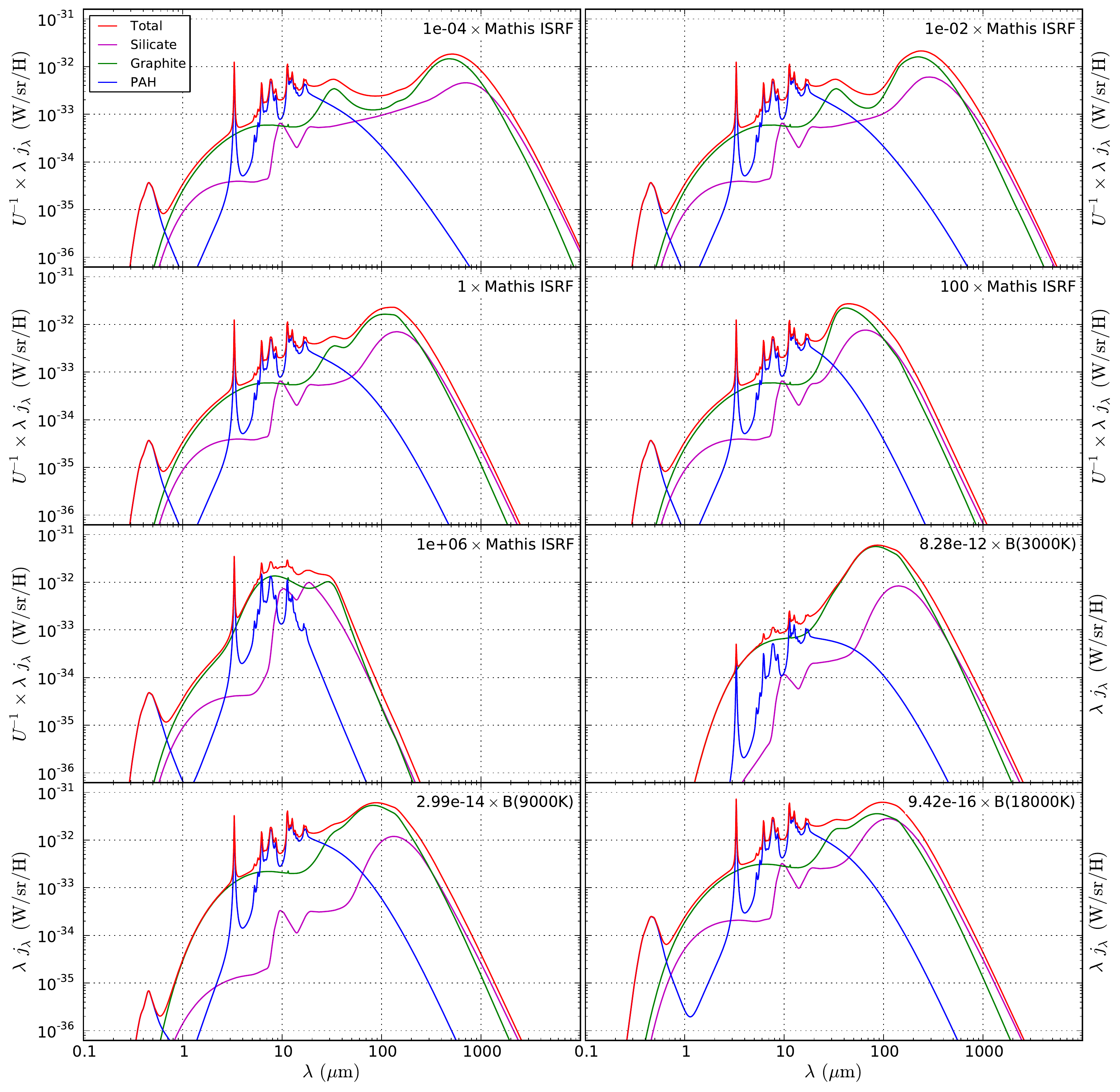}
  \caption{The reference solutions generated with the public version of DustEM (see Sect.\,\ref{sec:code-dustem}) using 3500 temperature bins and 250 iterations in the integral equation solver. The panels show the calculated dust emissivity for a selection of the input fields defined in Sect.\,\ref{sec:fields}. In each panel, the red curve represents the total emissivity and the other curves represent the portion of the emissivity for each grain type, silicate (magenta), graphite (green) and PAHs (blue). For the scaled Mathis input fields, the emissivity is divided by the input field strength $U$ to allow identical axis ranges for all plots.}
  \label{fig:dustem-solutions}
\end{figure*}

\begin{figure*}
  \centering
  \includegraphics[width=\textwidth]{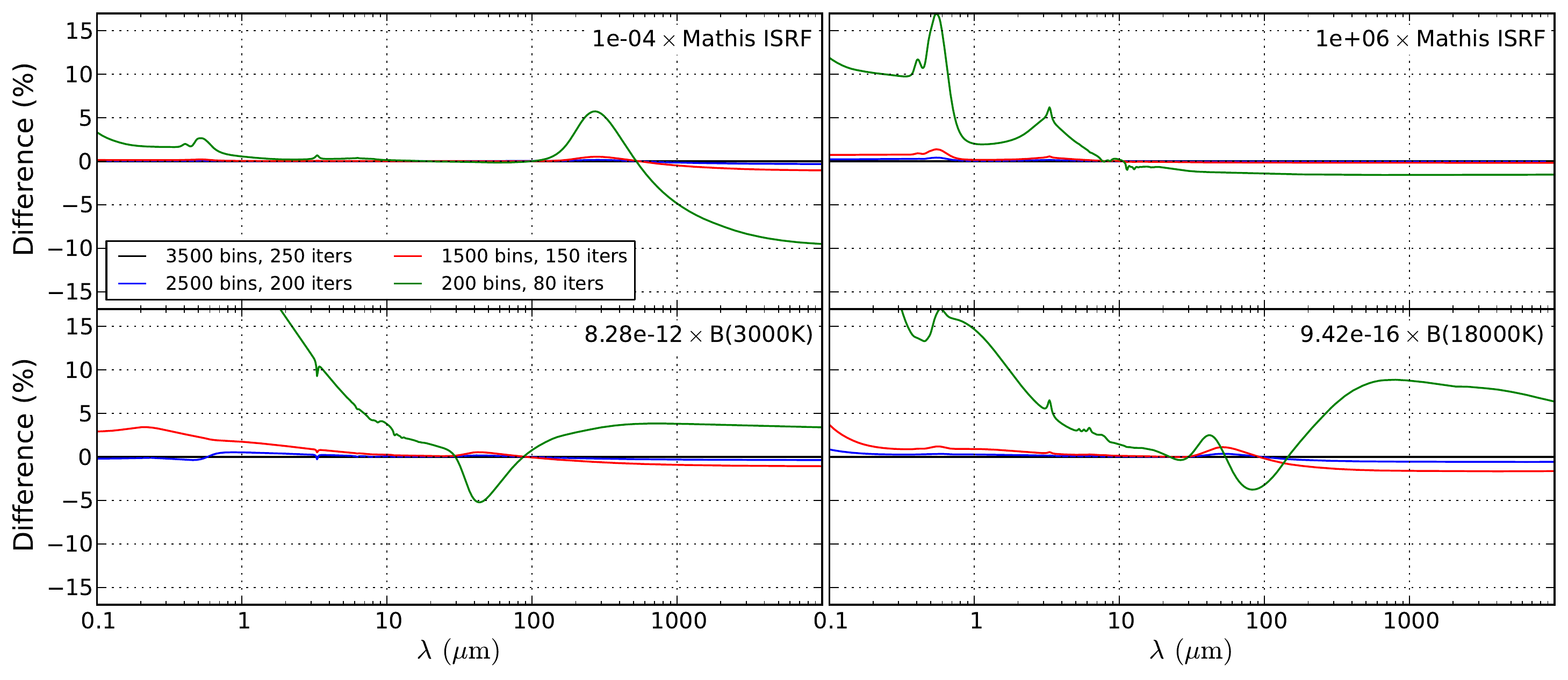}
  \caption{A comparison of DustEM solutions for the most extreme input fields defined in Sect.\,\ref{sec:fields}, calculated with a varying number of temperature bins and iterations in the DustEM integral equation solver. The solutions employed as a reference for our benchmark are calculated with 3500 temperature bins and 250 iterations; these solutions are represented in this figure by the zero-lines. The solutions calculated with the standard DustEM values of 200 temperature bins and 80 iterations are represented by the green curve. For these extreme fields, the standard solution deviates by up to 20\% (and even more for wavelengths shorter than 1 \micron). The solutions using 2500 temperature bins and 200 iterations (the blue curve) differ by less than 1\% from the reference solution, indicating numerical convergence at these parameter values. The contribution of each grain type separately has a similar convergence behavior (not shown).}
  \label{fig:dustem-convergence}
\end{figure*}

\begin{figure*}
  \centering
  \includegraphics[width=\textwidth]{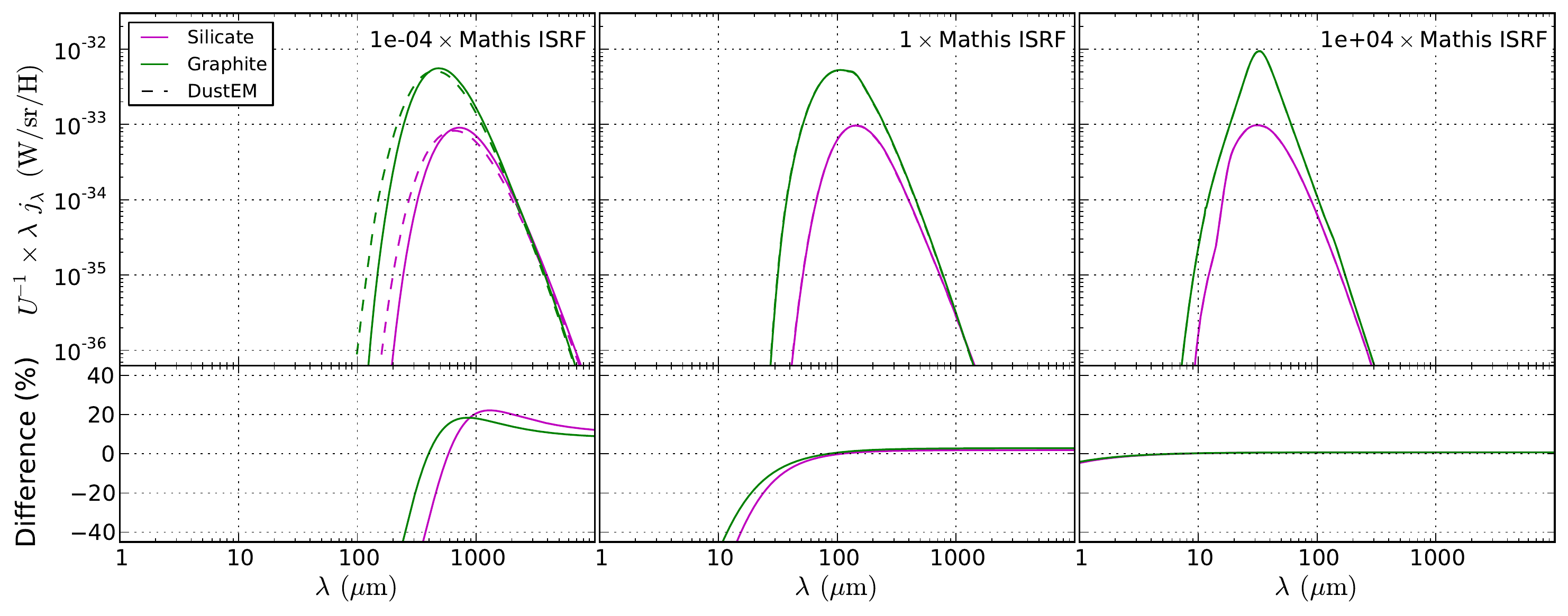}
  \caption{A comparison of the emissivities calculated by DustEM (using 3500 temperature bins and 250 iterations) for single-size, near-LTE grain populations to the corresponding equilibrium emissivities. The panels show the comparison for input fields ranging from extremely weak (left) to strong (right). The emissivity is divided by the input field strength $U$ to allow identical axis ranges for all plots. We used a dust mixture consisting of 0.05\,\micron\ silicate (magenta) or graphite (green) grains with a total dust mass per hydrogen atom of $10^{-30}$ kg/H for each grain type. The solid curves represent the emissivities calculated by one of our codes (SKIRT) under the assumption of LTE. The dashed curves represent the  solutions calculated by DustEM, without any LTE assumptions. The lower panels show the deviation of the equilibrium solutions from the corresponding full solutions. In a strong field, where we expect the grains to be in equilibrium, the solutions are indeed virtually identical. In a weaker field, the solutions differ since the grains are no longer completely in equilibrium.}
  \label{fig:dustem-equilibrium}
\end{figure*}

\begin{figure*}
  \centering
  \includegraphics[width=\textwidth]{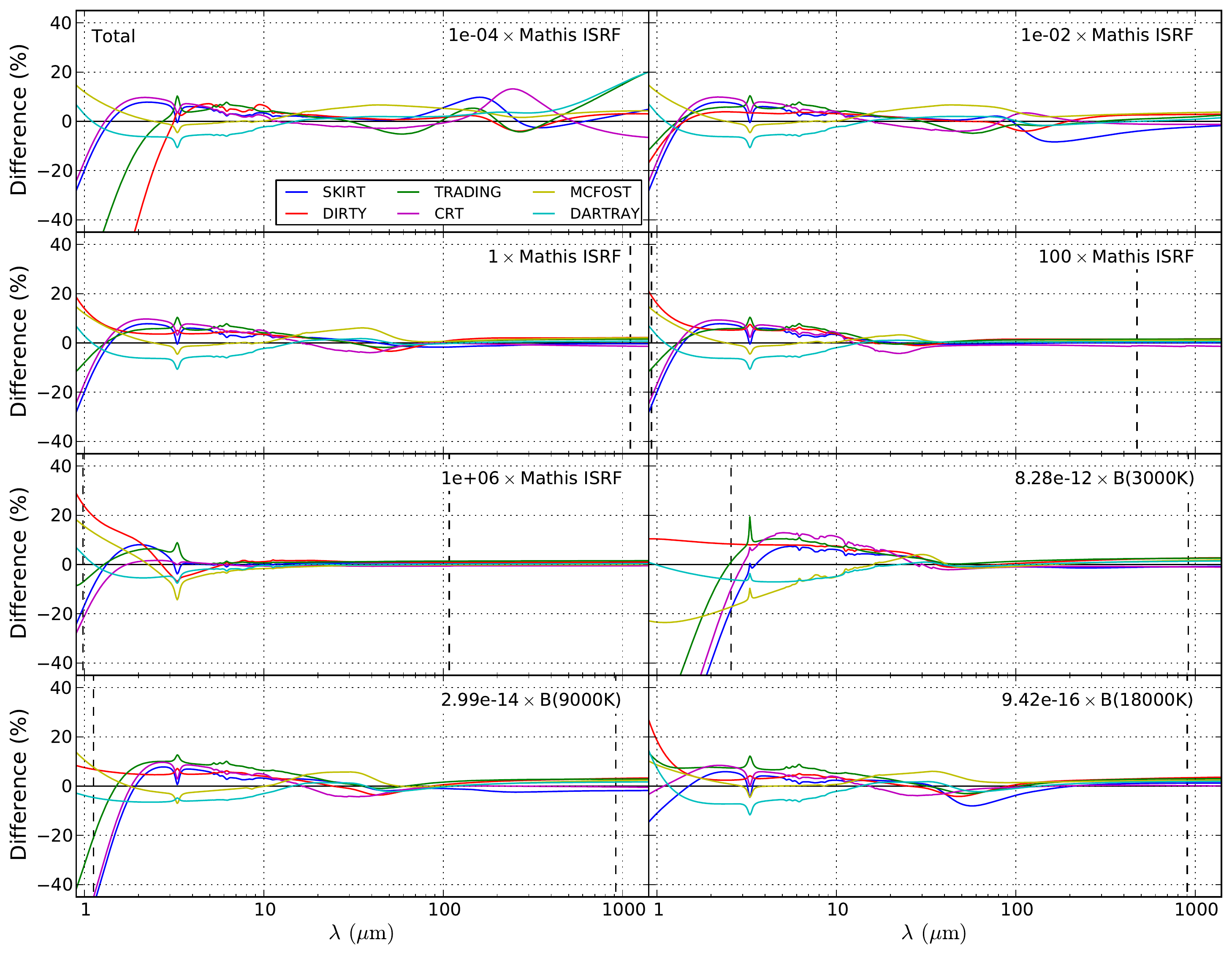}
  \caption{The relative differences between the total emissivities calculated by each of the codes participating in this benchmark and the corresponding reference solutions. The panels show the results for a selection of the input fields defined in Sect.\,\ref{sec:fields}. In each panel, the reference solution is represented by the zero-line. Positive percentages indicate results above the reference solution. The vertical dashed lines indicate where the reference solution becomes three orders of magnitude smaller than its peak value.}
  \label{fig:benchmark-total}
\end{figure*}

In the wavelength range $3\,\micron\le\lambda\le1000\,\micron$ all participating codes reproduce the total dust emissivity within 20\% of the reference solution for all input fields used in this benchmark (see Fig.\,\ref{fig:benchmark-total}). Excluding the weakest ($U\lesssim10^{-4}$) and the softest ($T\lesssim3000$\,K) fields, the correspondence in the same wavelength range is within 10\%. The larger relative deviations at wavelengths shorter than $3\,\micron$ are caused in part by the much lower absolute emissivity values in that range (two to three orders of magnitude below peak values; see Fig.\,\ref{fig:dustem-solutions} and the vertical dashed lines in Fig.\,\ref{fig:benchmark-total}).

The emissivities calculated for the silicate and graphite components (Figs.\,\ref{fig:benchmark-silicate} and \ref{fig:benchmark-graphite}) show a similar pattern, although the deviations in the individual components are sometimes slightly larger. The emissivities calculated for the PAH component (Fig.\,\ref{fig:benchmark-pah}) show the largest deviations. If we restrict the analysis to the wavelength range in which the emissivity of the reference solution is within three orders of magnitude of its peak value (indicated by the vertical dashed lines), the correspondence is still within 40\% for all codes and for all input fields, and often a lot better.

The larger discrepancies between the various codes for the PAH component can be traced to the fact that PAH molecules are, generally, substantially smaller than silicate and graphite grains; see the $a^\mathrm{max}$ values in Table \ref{tbl:baregrs}. As noted in Sect.\,\ref{sec:stoch-emission} and further discussed in Sect.\,\ref{sec:transition}, smaller grains are more likely to remain in the stochastic regime, requiring complex calculations that are more sensitive to differences in discretization (choice of grids) and concrete implementation (even if the same overall method is employed), as compared to equilibrium calculations.

Specifically, the PAH emissivities calculated by CRT (magenta curves in Fig.\,\ref{fig:benchmark-pah}) deviate from the other codes because, as described in Sect.\,\ref{sec:code-crt}, CRT implements the \citet{2001ApJ...551..807D} equations rather than Eqs.\,\ref{eq:coeff-heating} and \ref{eq:coeff-cooling} for calculating the heating and cooling transition rates. This method does not allow the emission of photons with an energy higher than the enthalpy content of the emitting grain \citep[see Eq. 56 in][]{2001ApJ...551..807D}. The upwards jumps in the emission spectrum appear when, going towards longer wavelengths (lower photon energy), a new enthalpy bin enters the emission calculation \citep[see also Figs.\,14 and 15 in][]{2001ApJ...551..807D}. These discontinuities appear only in the wavelength range where emission is largely from grains with very low enthalpy, for which the continuous cooling approximation is not valid, as described in Sect.\,\ref{sec:benchmark-solutions}.

Fig.\,\ref{fig:stoch-equil} shows the contributions to the total emissivity calculated by one of our codes (DIRTY) in the stochastic and equilibrium regimes for each of the three grain types, and for a number of input fields.\footnote{The precise form of these respective contributions varies between codes because of differences in transitioning from one regime to the other, but the general trend is similar.} While the silicate and graphite components show a significant equilibrium contribution for all fields, the PAHs remain in the stochastic regime for all but the strongest fields.

With respect to our conclusion at the end of Sect.\,\ref{sec:benchmark-solutions}, it is worth noting in Fig.\,\ref{fig:stoch-equil} that, at $\lambda=1000\,\micron$, the equilibrium contributions of the silicate and graphite components dominate the total stochastic contribution for all input fields. Only for the weakest fields ($U\approx10^{-4}$) does the stochastic contribution at that wavelength become a noticeable fraction of the total emission.

Considering the contributions of the different grain types to the total spectrum (Fig.\,\ref{fig:dustem-solutions}), it turns out that, for our dust model, the graphite emission dominates over most of the wavelength range for most input fields. The PAHs dominate in a very small region (their peak) and the silicates only at the longest wavelengths. Since the agreement between the different codes is significantly better for graphite and silicates than for the PAHs, our choice of dust mixture (unintentionally) benefits the global agreement between the different codes.

\subsection{Transition to equilibrium\label{sec:transition}}

As introduced in Sect.\,\ref{sec:stoch-emission}, and further elaborated upon in the code descriptions in Sect.\,\ref{sec:codes}, each code handles the transition from the stochastic to the equilibrium calculation regime in its own way. Figure\,\ref{fig:a-trans} shows the grain size $a_\mathrm{trans}$ for which the participating codes transition from the stochastic to the equilibrium calculation regime, for each grain type, and for each of the input fields defined in Sect.\,\ref{sec:fields}. While the details differ between codes, as a general trend, small grains (e.g. $a<0.01\,\micron$) are considered to be in equilibrium only when exposed to the strongest fields ($U>10^2$).

\begin{figure*}
  \centering
  \includegraphics[width=\textwidth]{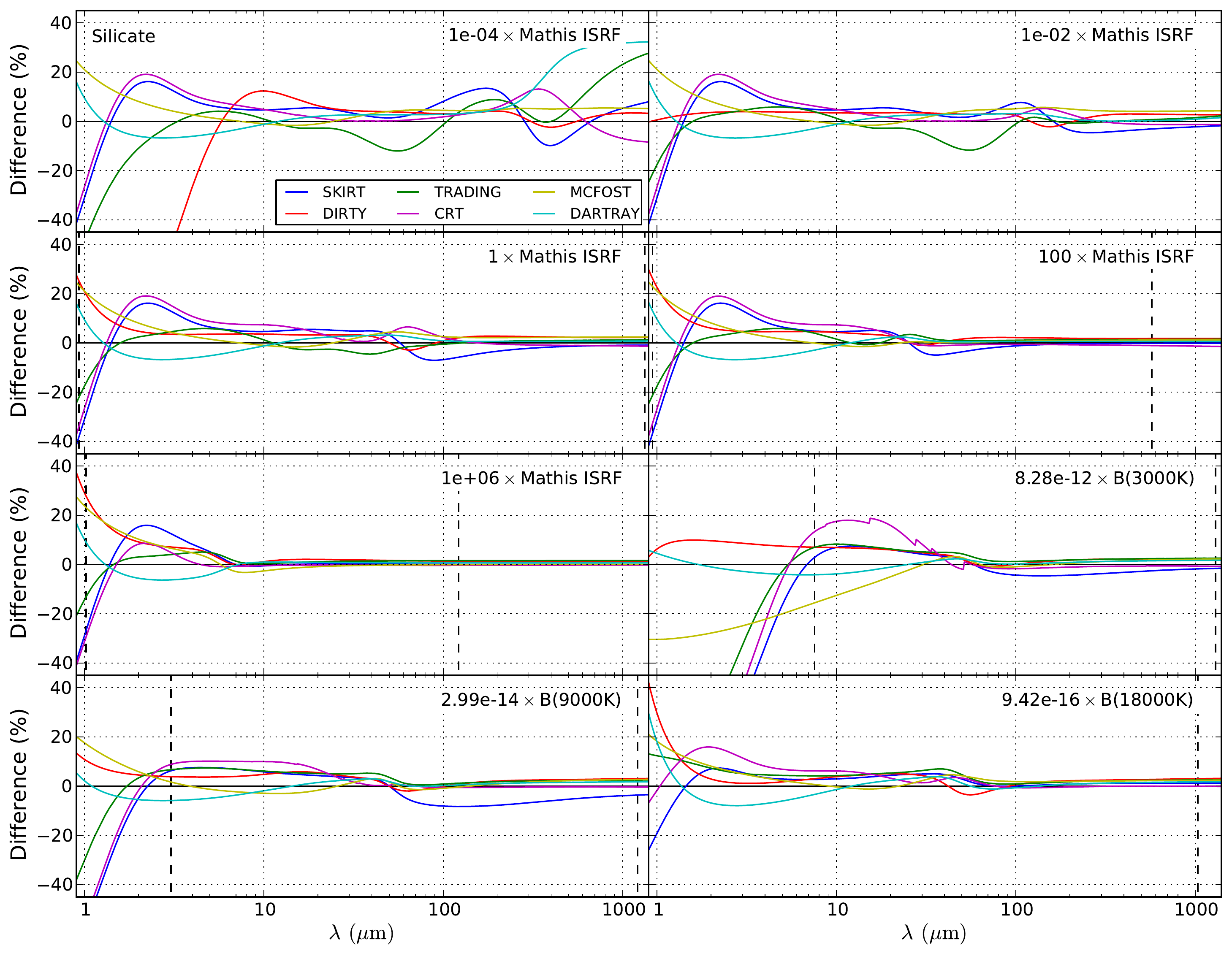}
  \caption{The relative differences between the emissivities of the silicate component calculated by each of the codes participating in this benchmark and the corresponding reference solutions. The panels show the results for a selection of the input fields defined in Sect.\,\ref{sec:fields}. In each panel, the reference solution is represented by the zero-line. Positive percentages indicate results above the reference solution. The vertical dashed lines indicate where the reference solution becomes three orders of magnitude smaller than its peak value.}
  \label{fig:benchmark-silicate}
\end{figure*}

\begin{figure*}
  \centering
  \includegraphics[width=\textwidth]{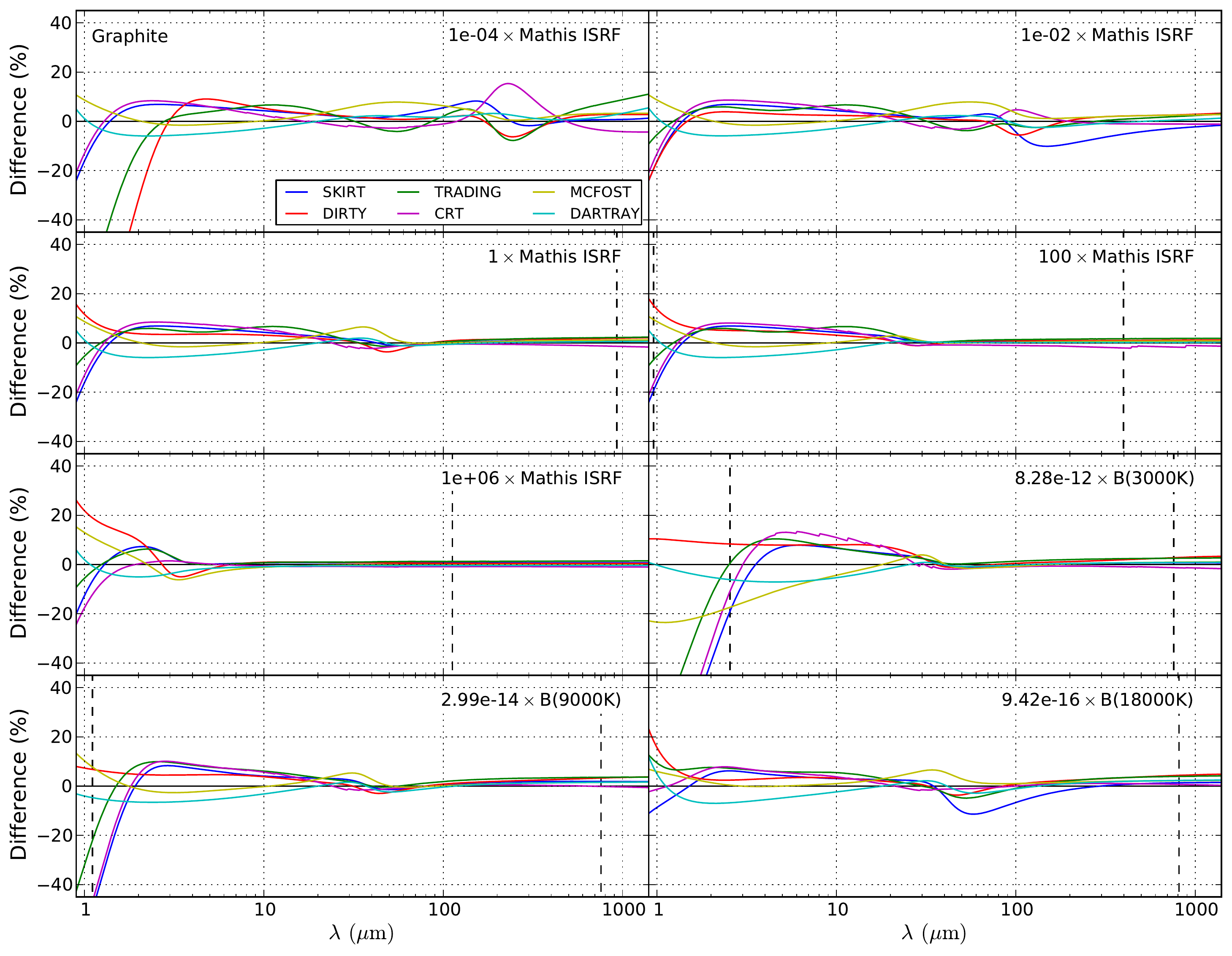}
  \caption{The relative differences between the emissivities of the graphite component calculated by each of the codes participating in this benchmark and the corresponding reference solutions. The panels show the results for a selection of the input fields defined in Sect.\,\ref{sec:fields}. In each panel, the reference solution is represented by the zero-line. Positive percentages indicate results above the reference solution. The vertical dashed lines indicate where the reference solution becomes three orders of magnitude smaller than its peak value.}
  \label{fig:benchmark-graphite}
\end{figure*}

\begin{figure*}
  \centering
  \includegraphics[width=\textwidth]{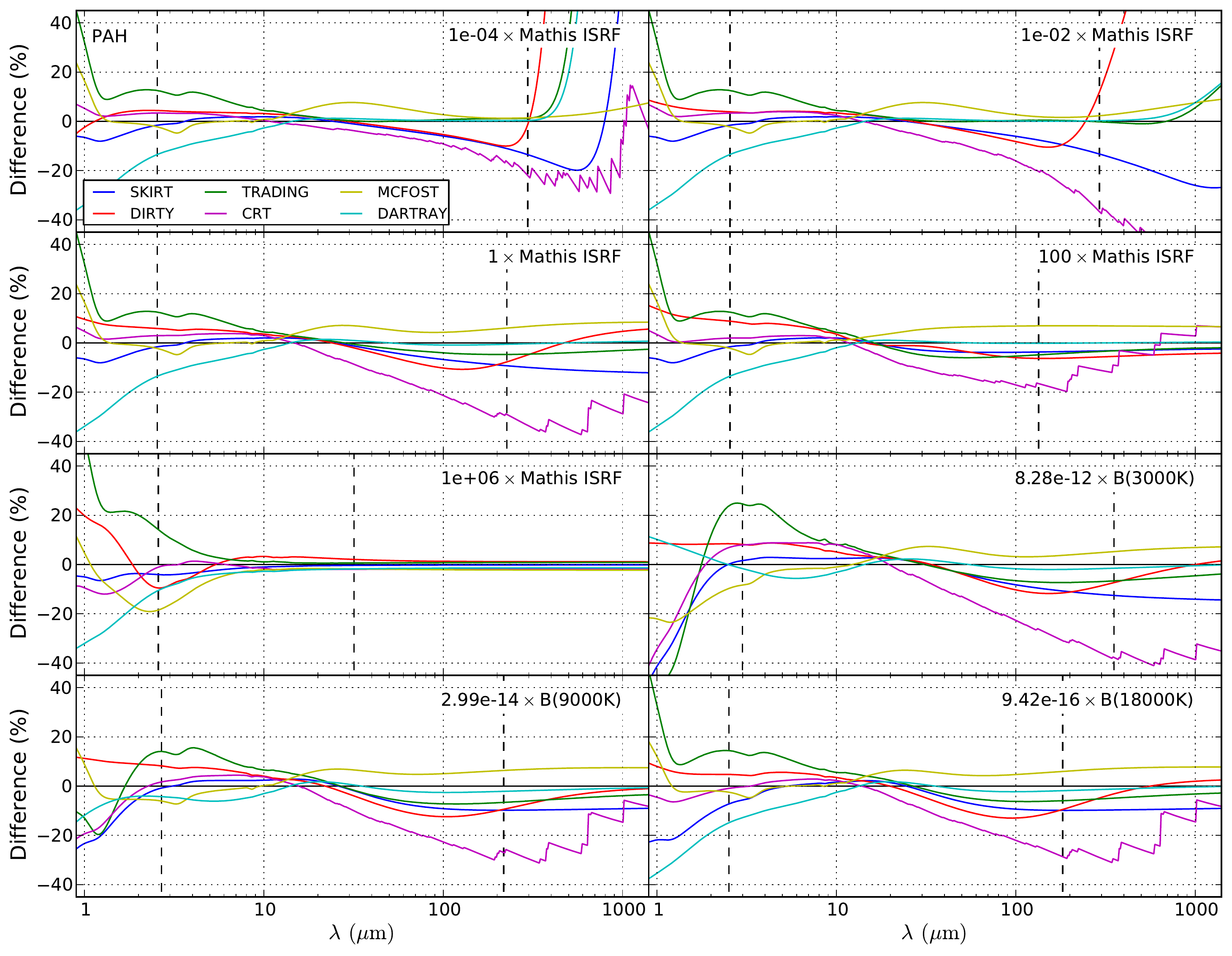}
  \caption{The relative differences between the emissivities of the PAH component calculated by each of the codes participating in this benchmark and the corresponding reference solutions. The panels show the results for a selection of the input fields defined in Sect.\,\ref{sec:fields}. In each panel, the reference solution is represented by the zero-line. Positive percentages indicate results above the reference solution. The vertical dashed lines indicate where the reference solution becomes three orders of magnitude smaller than its peak value.}
  \label{fig:benchmark-pah}
\end{figure*}

\begin{figure*}
  \centering
  \includegraphics[width=\textwidth]{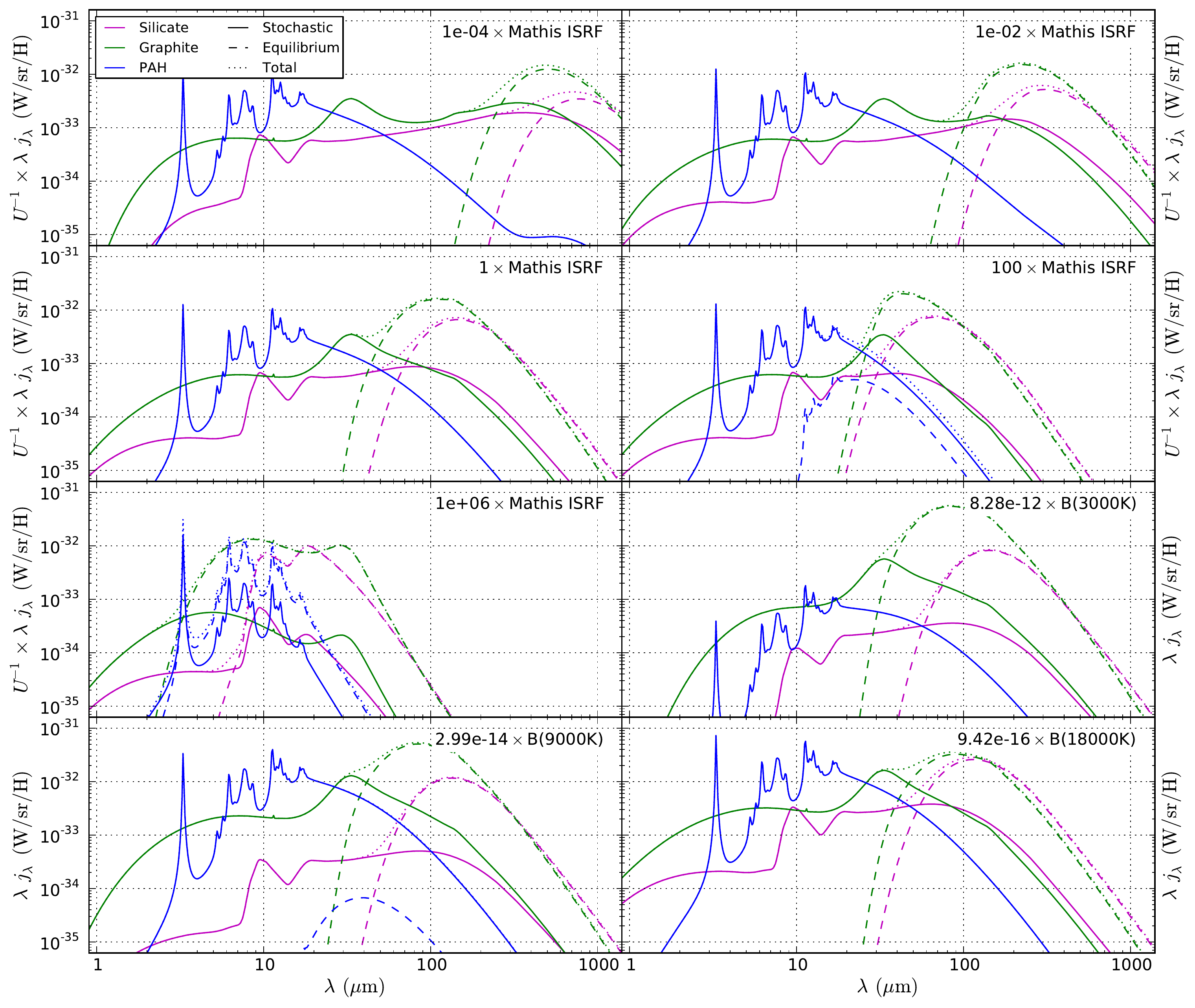}
  \caption{The contributions calculated in the stochastic (solid) and equilibrium (dashed) regimes to the total emissivity (dotted), for each of the grain types silicate (magenta), graphite (green) and PAHs (blue), by one of our codes (DIRTY). The panels show the results for a selection of the input fields defined in Sect.\,\ref{sec:fields}. For the scaled Mathis input fields, the emissivity is divided by the input field strength $U$ to allow identical axis ranges for all plots. The relative contributions for each regime may vary between codes because of the differences in the schemes to transition from one regime to the other.}
  \label{fig:stoch-equil}
\end{figure*}

\begin{figure*}
  \centering
  \includegraphics[width=\textwidth]{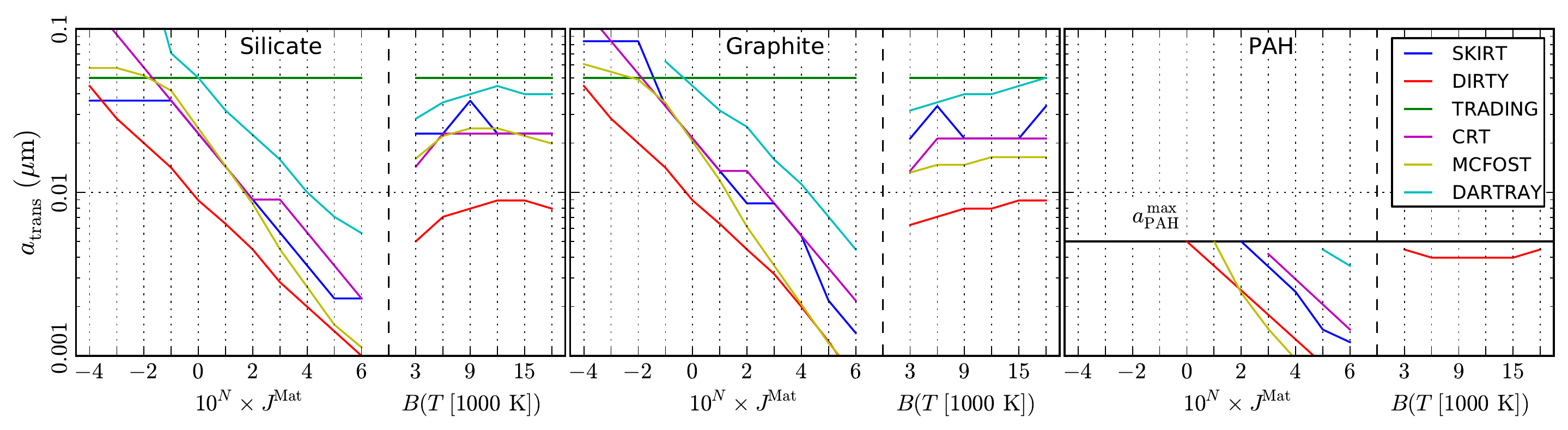}
  \caption{The grain size $a_\mathrm{trans}$ for which the participating codes transition from the stochastic to the equilibrium calculation regime. The panels show the smallest grain size that was considered to be in equilibrium for silicate (left), graphite (right) and PAH (right) grains, for each of the codes, across all of the input fields defined in Sect.\,\ref{sec:fields} (within each panel, the scaled Mathis ISRF fields to the left, and the diluted black body fields to the right).}
  \label{fig:a-trans}
\end{figure*}

Because most of the PAHs remain in the stochastic regime (Fig.\,\ref{fig:stoch-equil} and right panel of Fig.\,\ref{fig:a-trans}), the transition differences are most easily seen in the results for the silicate and graphite grains (Figs.\,\ref{fig:benchmark-silicate} and \ref{fig:benchmark-graphite}). As a general trend, the results converge for longer wavelengths, because the equilibrium emission dominating in that range is calculated in the same straightforward manner across all codes. The larger discrepancies are found at shorter wavelengths, where the stochastic regime dominates. Depending on the input field, the transition point shifts on the wavelength scale. Interestingly, for some fields, the discrepancies show extra ,,wiggles'' near the transition points, most likely caused by differences in handling the transition. This is particularly evident in the silicate emissivities for the scaled Mathis fields with strengths $U=10^{-2}$ to $10^2,$ (see the top half of Fig.\,\ref{fig:benchmark-silicate}).

The total emissivities are influenced by these transition differences mostly in the wavelength range just short-ward of the large submm emission peak, which is dominated by LTE emission from silicate and graphite grains (see Fig.\,\ref{fig:benchmark-total}). The position of this peak is determined by the temperatures of the dust grains, and thus depends on the input field. For the strongest field in our benchmark ($U=10^6$), the broad peak even overlaps the PAH features in the 3 to 30\,\micron\ wavelength range (see the third panel in the leftmost column of Fig.\,\ref{fig:benchmark-total}).

\subsection{Weak fields\label{sec:weak-fields}}

As discussed in Sect.\,\ref{sec:stoch-emission}, the transition size $a_\mathrm{trans}$ above which a grain can be considered to be in LTE depends on the radiation field to which the grain is exposed. For a given grain composition, $a_\mathrm{trans}$ tends to be higher for weaker fields because photon interactions are less frequent, which keeps larger grains in the stochastic regime. Consequently, the dust emission calculations are more complex for weaker fields, and especially for small grains in weaker fields. In addition to the computation time, this complexity affects the accuracy of the results, which explains why the largest discrepancies between the various solutions occur for the weakest field in the benchmark (see the top left panels in Figs.\,\ref{fig:benchmark-total} through \ref{fig:benchmark-pah}).

In fact, the weakest field in our benchmark ($U=10^{-4}$) may be unrealistically weak, since its peak intensity is below the peak of the CMB -- albeit in a different wavelength regime (UV versus mm wavelengths). To evaluate the effect of neglecting the CMB, we added the CMB to the $10^{-4}\times J^{\mathrm{Mat}}$ field, and used DustEM to recalculate the emissivity of our dust model exposed to this new input field. For wavelengths $\lambda\le100\,\micron$ the results are essentially identical to those shown in the top left panel of Fig.\,\ref{fig:dustem-solutions}. The submm peak is a notch higher and slightly shifted to longer wavelengths, causing an emissivity increase of about 35\% at $\lambda=1000\,\micron$. While this effect may not be negligible, it does not invalidate the benchmark test.

However, as argued in Sect.\,\ref{sec:benchmark-solutions}, our computations may no longer be physically founded for these weak fields, especially for small grains with internal energies comparable to those of the CMB photons. In conclusion, the $10^{-4}\times J^{\mathrm{Mat}}$ input field is properly benchmarking the various codes, but the calculations may be collectively incorrect because the continuous cooling approximation is inappropriate in this regime.

\subsection{Temperature discretization}

The participating codes implement various ways to discretize the grain temperature (or equivalently, the grain enthalpy), as described in Sect.\,\ref{sec:codes}. The different schemes are mostly driven by the aim to increase performance while preserving accuracy. Here we discuss the impact of respectively the minimum and maximum temperature values allowed in the grid.

DustEM, used to calculate our reference solutions, does not impose a lower temperature limit other than the zero point. Indeed, under the continuous cooling assumption, a small dust grain does not have to be in equilibrium with the CMB, and thus there seems to be no reason why the grain should not have, at a given moment in time, a temperature below 2.73\,K. In other words, we need to calculate the temperature probability distribution as usual. As argued in Sects.\,\ref{sec:benchmark-solutions} and \ref{sec:weak-fields}, this line of reasoning breaks down for small grains at very low energies, since the continuous cooling assumption no longer holds. 

This is why most codes participating in this benchmark do impose a lower temperature limit of 2.73\,K, or even, rather arbitrarily, 1, 2 or 3\,K. A limit of 2.73 K, for example, causes a bump in the PAH emission peaking at $\lambda\approx1060\,\micron$ (the peak CMB wavelength) because all the probabilities for lower temperatures are bunched together in the 2.73\,K temperature bin. This effect is to some extent responsible for the discrepancies between the codes and the reference solution seen in the top left panel of Fig.\,\ref{fig:benchmark-pah}, in the wavelength range to the right of the dashed line, where the absolute value of the emissivity has become small anyway. The effect is negligible for all but the weakest fields.

At the other end of the scale, all codes in this benchmark use the complete temperature range for which the dust properties are defined, i.e.\ up to 2500\,K for our dust model. As described in Sect.\,\ref{sec:benchmark-solutions}, this is well above the sublimation temperature of the dust material, although a fraction of the grains may survive at these temperatures for some time. Because the method used for this benchmark ignores dust grain destruction, we expect it to overestimate the emissivity for shorter wavelengths. To evaluate this effect, we reran the benchmark calculations with one of the codes (SKIRT) using a maximum grid temperature of 2250\,K instead of 2500\,K. As expected, the emissivities in the range $\lambda>3\,\micron$ are essentially unaffected by this change. At wavelengths shorter than $3\,\micron$, the total emissivity for the hardest black body input fields ($T\gtrsim15000$\,K) decreases noticeably (by about 30\% at 1 \micron), while there is no perceptible change down to 1 \micron\ for the softer fields or for the scaled Mathis fields. The silicate and graphite components behave similarly. Interestingly, the emissivity of the PAH component shows a noticeable decrease for all but the softest black body fields ($T\lesssim6000$\,K), and for all scaled Mathis fields. This is can be understood by recalling that the PAH particles are, on average, a lot smaller than those in the other components, so that they are more easily propelled to higher temperatures.

\subsection{Wavelength discretization}

The method described in Sect.\,\ref{sec:emission} for calculating dust emission involves wavelength discretization in several distinct areas. The optical dust properties are tabulated on some predefined wavelength grid. We also need to configure the wavelength range and sampling resolution for the input fields and for the output emissivity. And finally, the calculation of the cooling coefficients defined in Eq.\,\ref{eq:coeff-cooling} requires an integration of the optical properties over wavelength, which implies a grid as well. As long as proper interpolation procedures are in place, these wavelength grids do not need to be identical.

To keep matters simple for the benchmarks presented in this paper, Sect.\,\ref{sec:wavegrid} specifies the same wavelength grid for the optical properties and for the calculated emissivities, i.e.\ a logarithmic grid with 1201 points in the range $0.001\,\micron\leq\lambda\leq10000\,\micron$. In this section we discuss the impact of the resolution and of the lower and upper limits of the wavelength grid on the emissivity calculations. We reran the benchmark calculations with one of the codes (SKIRT) using some wavelength grid variations as reported below. SKIRT employs a single (configurable) wavelength grid for all aspects of the calculation. The optical dust properties are interpolated to this grid, and the grid is subsequently used for the input/output fields and for the integration to obtain the cooling coefficients. 

In practice, the lower wavelength limit is determined by the need to properly capture the input field in the calculation of the heating coefficients (Eq.\,\ref{eq:coeff-heating}) and the equilibrium temperature (Eq.\,\ref{eq:equilibrium}). For a scaled Mathis field, the lower limit can be increased to $0.09\,\micron$ (see Eq.\,\ref{eq:mathis}). For the hardest black body fields ($T\gtrsim12000$\,K), the limit should be lower. With a lower limit of $0.01\,\micron$ there is no noticeable difference in the calculated emissivities even for the hardest field in this benchmark ($T=18000$\,K)

The upper wavelength limit is mostly determined by the need to calculate the emissivity up to mm wavelengths. The upper limit also affects the calculation of the heating coefficients and the equilibrium temperature, but this effect is smaller because the absorption coefficients for the grain material are much lower at longer wavelengths. With an upper limit of $2000\,\micron$ instead of $10000\,\micron$ there is no noticeable difference in the calculated emissivities for any of the input fields. With an upper limit of $1000\,\micron$ the submm emissivity peak is overestimated by 20\% for the weakest scaled Mathis field ($U=10^{-4}$). The emissivity peak for the $U=10^{-3}$ field shows a similar but much smaller effect. For all other input fields there is no noticeable difference.

Finally, we reran the benchmark calculations for logarithmically distributed wavelength grids with successively lower resolution, always using a range of $0.01\,\micron\leq\lambda\leq2000\,\micron$. First of all, lowering the wavelength resolution affects the shape of the sharp PAH dominated emissivity peaks in the 3 to 30\,\micron\ wavelength range; if there is no wavelength point at the center of a peak, the peak can not be resolved. However, this does not affect the accuracy of the emissivity at other wavelength points (unless the resolution becomes too low, as described in what follows). Other than this peak resolution effect, using 601 wavelength points instead of 1201 does not noticeably influence the results for any of the input fields. Lowering the resolution to 301 wavelength points causes minor deviations in the calculated PAH emissivities for wavelengths $\lambda<3\,\micron$, which however do not noticeably affect the total emissivities down to $\lambda\ge1\,\micron$. With only 151 wavelength points the deviation in the total emissivities is still limited (a few percent at 1\,\micron), but the PAH features are now clearly under-resolved and rather smoothed out. This can be improved by concentrating more grid points in the wavelength range of the PAH features. For example, a specialty grid with a total of 151 points, 61 of which are concentrated in the 3 to 30\,\micron\ wavelength range, seems to provide an acceptable compromise.

\subsection{Grain size discretization}

The total calculation time for the emissivity of a dust population is roughly proportional to the number of grain size bins used to represent the population (see Sect.\,\ref{sec:mix-emission}). Therefore, some of the participating codes recompute the optical grain properties on an internally defined grain size grid rather than using the size bins tabulated in the dust model data. We used two of the codes (SKIRT and CRT) to investigate the effect of the number of grain size bins on the calculated emission. As mentioned in Sects.\,\ref{sec:code-skirt} and \ref{sec:code-crt}, these codes were configured with 15 size bins per grain type to calculate the benchmark results presented in Figs.\,\ref{fig:benchmark-total} through \ref{fig:benchmark-pah}.

The predominant effect of changing the number of grain size bins appears for wavelengths $\lambda<10\,\micron$ and grows larger for shorter wavelengths. This is to be expected; with a coarse grid the effective size of grains in the smallest bin is relatively large, and it is difficult to heat the grains to the high temperatures that are needed for emission at shorter wavelengths. Specifically, when the number of size bins per grain type is reduced from 15 to 10, the calculated silicate and graphite emissivities show substantial deviations from the reference solutions. The PAH emissivities are virtually unaffected, which is easily understood because their size range is much smaller. When the number of size bins per grain type is increased to 30, the calculated silicate and graphite emissivities in the wavelength range $1\,\micron<\lambda<10\,\micron$ approach the reference results. Again the PAH emissivities are much less affected because the dust model data has only 28 PAH size bins anyway.

A secondary effect occurs for longer wavelengths because the re-binning influences the heuristic for transitioning between the stochastic and equilibrium calculation regimes (Sect.\,\ref{sec:transition}). This effect seems to be somewhat random in nature, causing deviations that remain within the accuracy limits described in Sect.\,\ref{sec:eval-results}.

\subsection{Calculation time}

A typical 3D RT simulation calculates the dust emission spectra for a large number of dust cells. In cases where dust self-absorption is a relevant factor, this calculation is repeated for each iteration of the loop that self-consistently determines dust heating and re-emission (see e.g.\ Sect.\,\ref{sec:code-mcfost}). The time spent on calculating dust emission might thus become a significant or even dominant fraction of the total RT simulation time. The aim to reduce the dust emission calculation time has guided many of the choices in the implementations of the RT codes participating in this benchmark. Most fundamentally, all codes adopt the continuous cooling approximation. In addition, all codes select specific discretization schemes, most codes employ heuristics to transition between stochastic and equilibrium regimes, and some pre-compute field-independent data. Often these choices affect not only the calculation time, but also the accuracy of the results. In principle at least, the results can be made to match perfectly by increasing grid sizes and removing the heuristics, at the expense of calculation time.

It thus seems appropriate to consider calculation time when evaluating benchmark results. Unfortunately, it is not meaningful to compare the dust emission calculation times between the codes outside of the context of a RT simulation. For example, moving the relevant data for each dust cell from memory into the processor cache and back may represent a significant portion of the total calculation time, depending on the memory layout chosen by the RT code and depending on the architecture of the computer system. Consequently, a performance comparison would be more appropriately conducted as part of a RT benchmark.

Just to provide an order of magnitude, with the prescriptions provided in this paper, a code can calculate a few hundred dust emission spectra per second on a modern desktop computer. This means that the calculation for 5 million dust cells can be completed in a matter of hours rather than days.


\section{Conclusions\label{sec:conclusions}}

We defined an appropriate problem for benchmarking dust emissivity calculations in the context of RT simulations, specifically including the emission from stochastically heated dust grains (SHGs). The problem definition includes the optical and calorimetric material properties, and the grain size distributions, for a typical astronomical dust mixture with silicate, graphite and PAH components; a series of analytically defined radiation fields to which the dust population is to be exposed; and instructions for the desired output.

We summarized a popular method for calculating the emission from SHGs, with the intention to provide a self-contained guide for implementors of such functionality. The method is frequently used in RT codes because of its good performance and relative ease of implementation, although it assumes continuous cooling of the dust grains, which may be inaccurate in extreme environmental conditions. We then described the six RT codes participating in this benchmark effort, focusing on how their implementation of the SHG calculation differs, presenting relevant heuristics for accelerating the calculation, and studying the effects on the accuracy of the solutions. We also presented some practical hints with regards to the discretization of temperature and wavelength in the calculations. Most importantly, we processed the benchmark problem with each of the participating codes, and presented the results.

We reported in detail on the similarities and differences between the results from the participating codes and a reference solution. In the important wavelength range $3\,\micron\le\lambda\le1000\,\micron$ all participating codes reproduce the total dust emissivity within 20\% of the reference solution for all input fields used in this benchmark. Excluding the weakest and the softest input fields, the agreement in the same wavelength range is within 10\%. 

Our discussion offered hints on how RT codes could be enabled to properly calculate dust emission for a wider wavelength range. For example, when investigating systems with a lot of hot dust, such as circumstellar disks or accretion disks, it may be relevant to properly calculate dust emission for wavelengths shorter than $1\,\micron$. To accomplish this, RT codes will need to model environment-dependent destruction of dust grains, and adjust the grain size distribution used in the dust emission calculation accordingly.

In conclusion, this benchmark effort shows that the relevant modules in RT codes can and do produce fairly consistent results for the emissivity spectra of SHGs, which have a significant impact on the final result of a multi-wavelength RT simulation. We offer concrete, quantitative information on the level of (dis)agreement between RT codes, which will help inform the interpretation of RT simulation results that include SHG dust emission calculations of the type presented here. Specifically, this work paves the way for a more extensive benchmark effort focusing on the RT aspects of the various codes. And finally, we intend this work to serve as a reference for implementors of existing and new dust RT codes.


\begin{acknowledgements}
We thank the authors of DustEM \citep{2011AA...525A.103C} for making their code publicly available.
This work fits in the CHARM framework (Contemporary physical challenges in Heliospheric and AstRophysical Models), a phase VII Interuniversity Attraction Pole (IAP) programme organised by BELSPO, the BELgian federal Science Policy Office.
TL acknowledges the support from the Swedish National Space Board (SNSB).
GN is supported by the Leverhulme Trust research project grant RPG-2013-418.
MJ acknowledges the support of the Academy of Finland Grant No. 250741.
JS acknowledges support from the ANR (SEED ANR-11-CHEX-0007-01).
\end{acknowledgements}

\bibliographystyle{aa} 
\bibliography{shg}


\end{document}